\documentclass[a4paper,5p,fleqn]{cas-dc}
\usepackage[numbers]{natbib}

\usepackage{amsmath,amssymb}
\usepackage{graphicx}
\usepackage{textcomp}
\usepackage{multirow}
\usepackage{hyperref}
\usepackage{float}
\usepackage{xcolor,soul,framed} %color
\sethlcolor{yellow} 
\usepackage{pifont}

\usepackage{algpseudocode}

\usepackage{multicol}
\usepackage{subfig}
\usepackage{graphicx}

\usepackage[linesnumbered, ruled, vlined]{algorithm2e}
\usepackage{orcidlink}

\usepackage{nomencl} % Nomenclature package
\makenomenclature
\renewcommand*\nompreamble{\begin{multicols}{2}}
\renewcommand*\nompostamble{\end{multicols}}

\def\BibTeX{{\rm B\kern-.05em{\sc i\kern-.025em b}\kern-.08em
    T\kern-.1667em\lower.7ex\hbox{E}\kern-.125emX}}
    
\usepackage{threeparttable}
\usepackage{mathtools}

\newcolumntype{P}[1]{>{\centering\arraybackslash}p{#1}}

\def\tsc#1{\csdef{#1}{\textsc{\lowercase{#1}}\xspace}}
\tsc{WGM}
\tsc{QE}
\tsc{EP}
\tsc{PMS}
\tsc{BEC}
\tsc{DE}
\begin{document}

%\begin{multicols}{2}

% \let\WriteBookmarks\relax
% \def\floatpagepagefraction{1}
% \def\textpagefraction{.001}
\shorttitle{A Dual-Layer Image Encryption Framework Using Chaotic AES with Dynamic S-Boxes and Steganographic QR Codes}
\shortauthors{Bayesh et~al.}

\title [mode=title] {A Dual-Layer Image Encryption Framework Using Chaotic AES with Dynamic S-Boxes and Steganographic QR Codes}                      

\credit{Conceptualization of this study, Methodology, Software}

\address{Department of Computer Science and Engineering, Uttara University, Dhaka, Bangladesh.}

% \author{Dabbrata Das \orcidlink{0009-0008-0049-2048}}
% \ead{dasdabbrata@gmail.com}

% \author{Argho Deb Das \orcidlink{0009-0000-5756-2483}}
% \ead{atdeb727@gmail.com}

% \author{Farhan Sadaf \orcidlink{0009-0007-7550-7972}}
% \cormark[1]
% \ead{farhansadaf@cse.kuet.ac.bd}

\author{Md Rishadul Bayesh \orcidlink{0009-0008-1955-5766}}
\ead{rishad.cse18.kuet@gmail.com}

\author{Dabbrata Das \orcidlink{0009-0008-0049-2048}}
\cormark[1]
\ead{dasdabbrata@gmail.com}
 
\author{Md Ahadullah \orcidlink{0000-0002-8648-926X}}
\ead{ahadullah@uttara.ac.bd}

\cortext[cor1]{Corresponding author}

% \fntext[fn1]{Data: https://www.kaggle.com/datasets/soumikrakshit/nyu-depth-v2}
\fntext[fn2]{Code: https://github.com/rishad086/Image-Encryption}

\begin{abstract}
This paper presents a robust image encryption and key distribution framework that integrates an enhanced AES-128 algorithm with chaos theory and advanced steganographic techniques for dual-layer security. The encryption engine features a dynamic ShiftRows operation controlled by a logistic map, variable S-boxes generated from a two-dimensional Hénon map for substitution and key expansion, and feedback chaining with post-encryption XOR diffusion to improve confusion, diffusion, and key sensitivity. To address secure key delivery, the scheme introduces dual-key distribution via steganographically modified QR codes. A static key and an AES-encrypted dynamic session key are embedded with a covert hint message using least significant bit (LSB) steganography. This design ensures the dynamic key can only be decrypted after reconstructing the static key from the hidden message, offering multi-factor protection against interception. Experimental results demonstrate the framework outperforms existing chaos-based and hybrid AES methods, achieving near-ideal entropy (~7.997), minimal pixel correlation, and strong differential resistance with NPCR (>99.6\%) and UACI (~50.1\%). Encrypted images show uniform histograms and robustness against noise and data loss. The framework offers a scalable, secure solution for sensitive image transmission in applications such as surveillance, medical imaging, and digital forensics, bridging the gap between cryptographic strength and safe key distribution.

\end{abstract}

\maketitle

% \begin{table*}[!ht]   

% \nomenclature{$AES$}{Advanced Encryption Standard}
% \nomenclature{$S\text{-}Box$}{Substitution Box}
% \nomenclature{$IoT$}{Internet of Things}
% \nomenclature{$CBC$}{Cipher Block Chaining}
% \nomenclature{$ECB$}{Electronic Code Book}
% \nomenclature{$QR$}{Quick Response}
% \nomenclature{$PSNR$}{Peak Signal-to-Noise Ratio} 
% \nomenclature{$SSIM$}{Structural Similarity Index Measure}
% \nomenclature{$NPCR$}{Number of Pixel Change Rate}
% \nomenclature{$UACI$}{Unified Average Changing Intensity}
% \nomenclature{$ECC$}{ Elliptic Curve Cryptography}
% \nomenclature{$RSA$}{Rivest, Shamir, Adleman}
% \nomenclature{$CTR$}{Counter mode}
% \nomenclature{$CCM$}{Counter with CBC-MAC}
% \nomenclature{$GCM$}{Galois/Counter Mode}
% \nomenclature{$MAES$}{Modified AES}
% \nomenclature{$IV$}{Initialization Vector}
% \nomenclature{$QR$}{Quick Response}

% \begin{framed}
% \printnomenclature
% \end{framed}

% \end{table*}

\section{Introduction}
 \label{sec:intro}
With the rapid advancement of digital technologies and the exponential growth of image-centric data in sectors such as healthcare, surveillance, smart environments, and the Internet of Things (IoT), ensuring the confidentiality and integrity of visual information has become a critical concern for modern cybersecurity research~\cite{al2021survey, ahmed2020iot}. Unlike textual data, images possess high redundancy, strong spatial correlation among neighboring pixels, and large file sizes—all of which render conventional text-oriented encryption schemes, such as the Advanced Encryption Standard (AES) \cite{aes}, less effective in their direct form. Although AES offers strong cryptographic protection, when applied to images using its standard configuration (static S-box and fixed block modes), it often leaves behind detectable patterns in the encrypted output. These residual patterns pose significant risks, potentially exposing the data to cryptanalytic attacks, such as differential and statistical analyses.

To mitigate these shortcomings and bolster the security of image transmission over open or untrusted networks, this paper introduces a comprehensive and innovative image encryption scheme. The proposed framework enhances the AES algorithm \cite{aes} with chaos theory and introduces an efficient and visually meaningful secure transmission method for encryption keys. The system is designed around two core components: a chaos-augmented AES encryption model and a secure dual-key sharing mechanism using customized QR code transformation.

In the first phase of encryption, the image is encrypted using a dynamically generated key, providing high randomness and uniqueness per session. This dynamic key is then securely encrypted using a static key, creating a second layer of protection. To facilitate safe transmission of both keys to the receiver without exposing them directly, a QR code is generated, embedding three critical elements: (1) the encrypted dynamic key, (2) a slightly altered version of the static key, and (3) an encoded message that conveys how the static key was modified. The encoded message is created using a custom encoder and is essential for key restoration. This QR code is then visually customized—for example, rendered into an all-white image or modified in such a way that it disguises its true nature—allowing it to travel inconspicuously over open channels alongside the encrypted image.

At the core of the encryption process, a significantly modified AES algorithm is employed, incorporating a range of novel cryptographic enhancements. One of the central improvements is the use of a variable S-box, dynamically generated through a fast chaotic map. Unlike the fixed S-box~\cite{ssbox} in conventional AES, this evolving S-box changes at each encryption round, adding substantial key sensitivity and nonlinearity. This dynamic behavior ensures that the same plaintext image, encrypted under slightly different conditions, yields entirely different cipher outputs, thereby neutralizing various forms of attack, including differential and linear cryptanalysis.

To further obscure pixel-level correlations, the image is first masked using a logistic map-based XOR operation, which introduces confusion before encryption. Subsequently, the image pixels are reshaped and permuted using chaotic sequences to disrupt their spatial coherence, thereby achieving strong diffusion. The AES encryption process \cite{aes} itself is enhanced through block shuffling and shifting mechanisms, and integrates both 1D and 2D S-box configurations—a significant structural deviation from standard AES \cite{aes}. Additionally, a descending harmony chaotic map is introduced to generate more unpredictable values for cipher components.

Another noteworthy feature is the implementation of a hybrid block cipher mode, merging the Cipher Block Chaining (CBC) \cite{cbc_ecb} and Electronic Codebook (ECB) \cite{cbc_ecb} modes using XOR operations, but excluding the use of an initialization vector (IV). This hybrid mode benefits from the error propagation strengths of CBC and the speed of ECB, while minimizing their respective drawbacks.

The receiver-side decryption process precisely mirrors each of the encryption stages. Upon receiving the encrypted image and the visually disguised QR code, the receiver restores the original QR code, decodes the embedded message to reconstruct the static key, decrypts the dynamic key, and finally uses it to decrypt the image. Each stage is reversed systematically, ensuring accurate image recovery.

Extensive simulations and experiments validate the efficacy of the proposed method. To assess the fidelity and visual quality of decrypted images, widely accepted evaluation metrics such as Peak Signal-to-Noise Ratio (PSNR) and Structural Similarity Index Measure (SSIM) are employed~\cite{hore2010image}. Additionally, information entropy analysis, pixel correlation measurements, and histogram uniformity evaluations are utilized to verify the proposed encryption scheme's robustness against common cryptanalytic attacks~\cite{shannon1949communication, chen2004new}. The system also demonstrates resilience to post-encryption operations such as pixel shuffling, provided the correct decryption parameters are preserved.

In summary, the proposed image encryption framework offers a multi-layered and resilient security architecture by combining dynamic cryptographic components, chaotic systems, and advanced key transmission strategies. The fusion of visual obfuscation techniques with chaos-enhanced AES encryption provides a powerful solution for image security in modern digital communication environments, especially where data privacy is paramount.

\begin{table}[!ht]
\caption{\textbf{Nomenclature:} Abbreviations and their definitions.}
\begin{framed}
\centering
\begin{tabular}{ll}
\textbf{Abbreviation} & \textbf{Definition} \\
\hline
AES & Advanced Encryption Standard \\
S-Box & Substitution Box \\
IoT & Internet of Things \\
CBC & Cipher Block Chaining \\
ECB & Electronic Code Book \\
QR & Quick Response \\
PSNR & Peak Signal-to-Noise Ratio \\
SSIM & Structural Similarity Index Measure \\
NPCR & Number of Pixel Change Rate \\
UACI & Unified Average Changing Intensity \\
ECC & Elliptic Curve Cryptography \\
RSA & Rivest, Shamir, Adleman \\
CTR & Counter mode \\
CCM & Counter with CBC-MAC \\
GCM & Galois/Counter Mode \\
MAES & Modified AES \\
IV & Initialization Vector \\
\end{tabular}
\end{framed}
\end{table}

\section{Related Works} \label{sec:related works}
Numerous techniques have been developed to strengthen image encryption, particularly to address vulnerabilities in traditional cryptographic models like AES when applied to large and high-correlation image data. These approaches broadly fall into categories such as chaos-enhanced AES models, hybrid encryption frameworks, algorithmic performance evaluations, and lightweight or transform-based schemes. This section presents a detailed comparative review of representative works within each category, along with their limitations, future scope, and reported results.

\subsection{Techniques for Image Encryption Based on Chaos}

Recent advancements in image encryption have increasingly leveraged chaos theory due to its fundamental properties for example, susceptibility to beginning conditions, pseudo-randomness, and ergodicity. These characteristics allow chaotic systems to attain significant degrees of diffusion and confusion, making them excellent candidates for protecting digital images from statistical and differential attacks.

A notable example of standalone chaos-based encryption is presented by Yang et al. \cite{lorenz_chaos}, who proposed a bio-inspired approach combining a simplified fractional-order Lorenz chaotic system \cite{lorenz_chaotic_system} with DNA coding \cite{dna_coding}. The encryption process utilizes chaotic dynamics for pixel permutation and DNA-based substitution to enhance resistance against cryptanalytic attacks. Experimental results on the Peppers image demonstrated high performance, with an NPCR of 99.60\% and a UACI of 33.62\%, reflecting excellent sensitivity and randomness.

Similarly, Neamah \cite{7d_chaotic} introduced a two-stage encryption scheme based on a 7D hyperchaotic system. The design applies chaotic permutation followed by diffusion through a Pascal matrix \cite{pascal_matrix}, effectively spreading pixel intensity variations and strengthening the cipher's robustness. This model also reported strong security metrics on the Peppers image, achieving a UACI of 33.42\% and an NPCR of 99.58\%.

While these chaos-only models offer strong performance, other research has explored integrating chaotic systems with traditional symmetric encryption standards such as AES \cite{aes} to further enhance security. For instance, Hadj Brahim et al. \cite{hadj2024image} proposed a chaos-enhanced AES variant that dynamically generates S-boxes using a hyperchaotic system. Each 128-bit image block is encrypted with a uniquely generated S-box derived from ascending sequences, where the output of one sequence serves as the seed for the next. The model showed excellent diffusion performance, achieving a UACI of 33.44\% and an NPCR of 99.59\% when tested on the widely used Lena image---indicating a high level of sensitivity to pixel changes and strong resistance to differential attacks.

Ali et al. \cite{light_weight_aes} proposed another variation—a lightweight AES model optimized for real-time performance using Lorenz chaotic functions for dynamic key generation. This approach replaces the standard MixColumns transformation with bit-level circular permutation and integrates dynamic S-boxes to increase complexity. The model achieved rapid encryption and decryption times of 0.20558 and 0.28773 seconds, respectively. However, the S-box generation process remained relatively time-consuming, requiring approximately 1.5555 seconds.
Despite these efficiencies, the scheme retained a static key across encryption rounds, which may reduce resilience if the key is compromised.

Despite their promising results, these chaos-based and hybrid models present some limitations. The static-key dependency in Ali et al.’s model \cite{light_weight_aes} poses a vulnerability in scenarios where key exposure is possible. Additionally, Hadj Brahim et al.’s dynamic S-box approach \cite{hadj2024image} exhibited limited resistance to partial data loss due to insufficient pixel dispersion \cite{sbox}. Moreover, most of these encryption schemes have not yet been rigorously validated on resource-constrained platforms, such as those used in IoT or embedded environments.

Future studies ought to focus on filling in these gaps by incorporating round-wise key evolution, optimizing pixel scrambling strategies for better data loss resilience, and ensuring real-time feasibility on lightweight hardware. Furthermore, hybrid frameworks that fuse chaos theory with cryptographic standards like AES could offer a balanced trade-off between complexity, performance, and security in practical deployments.

\subsection{Hybrid Cryptographic Frameworks (AES + ECC/RSA)}
 A number of recent hybrid encryption models have explored the integration of traditional cryptographic algorithms with modern lightweight techniques to address security and performance challenges in image encryption. Benssalah et al. \cite{elliptic_curve_cryptography} proposed one such approach combines AES with Elliptic Curve Cryptography (ECC) \cite{ecc}, specifically tailored for medical imaging applications. This model enhances efficiency by removing the AES MixColumns step and replacing it with simpler column shifts, while ECC \cite{ecc} operations are offloaded to dedicated hardware for improved performance. Experimental results show that this approach \cite{elliptic_curve_cryptography} achieves a Unified Average Changing Intensity (UACI)\cite{npcr_uachi} of 30.3842\% on Lena image data, demonstrating good encryption strength and randomness.

Gafsi et al. \cite{rsa_chaotic} proposed another study introducing a hybrid system that combines RSA encryption \cite{rsa} with chaos-based confusion and diffusion. In this framework, RSA handles the confusion phase—ensuring key security—while chaotic maps are employed for the diffusion phase, enhancing pixel-level randomness in the encrypted image. These methods emphasize security through complexity and unpredictability. However, despite their theoretical robustness, practical implementation reveals several limitations. Hardware dependence restricts portability, and high computational overhead hampers real-time applicability, especially in resource-constrained environments. Moreover, the lack of a secure and user-friendly key-sharing mechanism limits their deployment potential.

Lightweight cryptographic alternatives to ECC \cite{ecc} and RSA \cite{rsa} can improve performance on constrained devices. Visual key-sharing methods, such as QR code encoding, offer a more secure and user-friendly approach to key distribution. Asynchronous hybrid models with multi-platform compatibility enable broader and more practical deployment.

\begin{table*}[!htbp, align=\flushleft, width=\textwidth]
    \caption{An overview of significant contributions to image encryption}
    \centering
    \renewcommand{\arraystretch}{1.25}
    \footnotesize
    \begin{threeparttable}
    \begin{tabular}{P{2.4cm} P{2.2cm} P{2.6cm} P{3.2cm} P{2.0cm} P{2.3cm}}
        \hline \hline
        \textbf{Reference} & \textbf{Type of Analysis} & \textbf{Model} & \textbf{Technique} & \textbf{Image} & \textbf{Performance} \\
        \hline
        Hadj Brahim et al. \cite{hadj2024image} & Chaos-enhanced AES & AES with variable S-box & Hyperchaotic system generates round-wise S-box using ascending sequences & Lena & NPCR: 99.5911\%, UACI: 33.4417\% \\
        \hline
        Ali et al. \cite{light_weight_aes} & Lightweight AES variant & Chaotic AES with Lorenz function & Dynamic key generation, circular bit permutation, dual S-boxes & Not specified & Enc: 0.20558s, Dec: 0.28773s, S-box Gen: 1.5555s \\
        \hline
        Benssalah et al. \cite{elliptic_curve_cryptography} & Hybrid (AES + ECC) & AES + ECC co-design & Column-shifted AES + hardware-accelerated ECC & Lena & UACI: 30.3842\% \\
        \hline
        Gafsi et al. \cite{rsa_chaotic} & Hybrid (RSA + Chaos) & RSA + Chaotic Diffusion & RSA for confusion; chaotic maps for pixel diffusion & Not specified & Theoretical strengths; high overhead; limited portability \\
        \hline
        Aladdin et al. \cite{comparative_analysis_cha_twofish} & Algorithm Evaluation & AES, Blowfish, Twofish, Salsa20, ChaCha20 & Performance analysis across ciphers & Various images & ChaCha20: Enc 0.0116ms, Dec 0.0194ms \\
        \hline
        Chen et al. \cite{all_aes_algo} & AES Mode Comparison & AES-128 ECB, CBC, etc. & Gini impurity-based evaluation of encryption modes & Not specified & ECB: 2.60609ms, CBC: 2.72313ms \\
        \hline
        Rehman et al. \cite{dwt} & Transform-based AES & KSP-DWT-IET & Key substitution with DWT, single-round encryption & Lena & UACI: 33.6\%, NPCR: 33.6\% (RGB channels) \\
        \hline
        Yang et al. \cite{lorenz_chaos} & Chaos + Bio-Inspired & Simplified Lorenz System + DNA Coding & Fractional Lorenz chaos with DNA-based encryption & Peppers & UACI: 33.62\%, NPCR: 99.60\% (Avg.) \\
        \hline
        Neamah \cite{7d_chaotic} & Chaos-based Image Encryption & 7D Hyperchaotic System & Two-stage encryption via permutation and Pascal matrix diffusion & Peppers & UACI: 33.42\%, NPCR: 99.58\% \\
        \hline
    \end{tabular}
    \end{threeparttable}
    \label{tab:image_encryption_works}
\end{table*}

\subsection{Performance Evaluation of AES and Other Symmetric Algorithms}
A comparison of symmetric encryption methods algorithms— encryption/decryption speed and throughput across various image types. Aladdin et al. \cite{comparative_analysis_cha_twofish} perform a comparative analysis of AES \cite{aes}, Blowfish \cite{blowfish}, Twofish \cite{twofish}, Salsa20 \cite{salsa20}, and ChaCha20 \cite{chacha20} for image encryption. Among these, ChaCha20 \cite{chacha20} demonstrated superior performance, with average encryption and decryption times of 0.0116 ms and 0.0194 ms, respectively, highlighting its efficiency for lightweight and fast image encryption tasks. These findings underscore ChaCha20’s suitability for applications demanding low latency and high throughput.

Another study proposed by Chen et al. \cite{all_aes_algo} explored the impact of different AES modes, including ECB \cite{cbc_ecb}, CBC \cite{cbc_ecb}, CTR \cite{ctr}, CCM \cite{ccm_gcm}, and GCM \cite{ccm_gcm}—on encryption performance. The evaluation employed a novel Gini impurity-based metric to assess the randomness and distribution of encrypted data. AES-128 in ECB and CBC modes \cite{cbc_ecb} reported encryption times of 2.60609 ms and 2.72313 ms, respectively. While the study provided insights into operational efficiency, its primary focus remained on computational aspects rather than comprehensive security evaluations.

Although these works demonstrate performance benefits, they often overlook critical visual security parameters such as entropy, correlation, and differential attack resistance (e.g., NPCR and UACI). Additionally, the effectiveness of encryption may vary significantly with image content, reducing the generalizability of the results. Incorporating robust visual security metrics and evaluating ciphers under dynamic conditions like real-time streaming and frame-wise encryption could offer a more holistic assessment of their applicability in practical scenarios.

\subsection{Efficient Image Encryption with Transform and Substitution Techniques}
Rehman et al. \cite{dwt} present the KSP-DWT-IET method \cite{dwt}, which integrates key substitution with the Discrete Wavelet Transform (DWT) to achieve low-latency image encryption in a single round, leveraging frequency-domain compression and novel weighted key schemes. Experimental results on the Lena image demonstrate strong statistical performance, with UACI \cite{npcr_uachi} values of 33.6487\% (R), 33.6145\% (G), and 33.6554\% (B), and NPCR \cite{npcr_uachi} values of 33.6474\% (R), 33.6512\% (G), and 33.6651\% (B). However, the algorithm has not been evaluated for high-resolution images or multimedia content such as video, and its resilience against advanced cryptanalytic techniques over time remains unexamined. Future research should focus on extending this approach to multi-frame video encryption, incorporating chaotic DWT filters \cite{chaotic_dwt} or hybrid transforms to enhance diffusion properties, and assessing its viability in live surveillance systems are examples of real-time applications.\\

To address the challenges identified in prior work, this paper presents a comprehensive image encryption framework with the following key contributions:

\begin{itemize}
\item Dual-Key secure transmission framework: A static key securely transmits a dynamic key used for modified AES image encryption.
\item Steganographic QR code embedding with visual customization: Embeds a modified static key, encrypted dynamic key, and an unscannable key-alteration message for secure and covert transmission..
\item Adaptive key retrieval via encoded message: Decoder reconstructs the original static key from the altered version using encoded message.
\item Chaos-enhanced AES with dynamic S-box substitution: Replaces standard AES S-box with round-varying ones from a fast chaotic map to enhance nonlinearity and security.
\item Pre-AES chaotic pixel masking and permutation: Pixels are XOR-masked and permuted via logistic map before AES encryption to increase entropy.
\item Post-encryption pixel shuffling for enhanced robustness: Additional pixel rearrangement after AES strengthens resistance to tampering while preserving decryptability.
\end{itemize}

\section{Methodology}

\subsection{Preliminaries}

\subsubsection{Asymmetric cryptosystem}

\begin{figure}[htbp]
    \centering
    \includegraphics[width=0.9\linewidth]{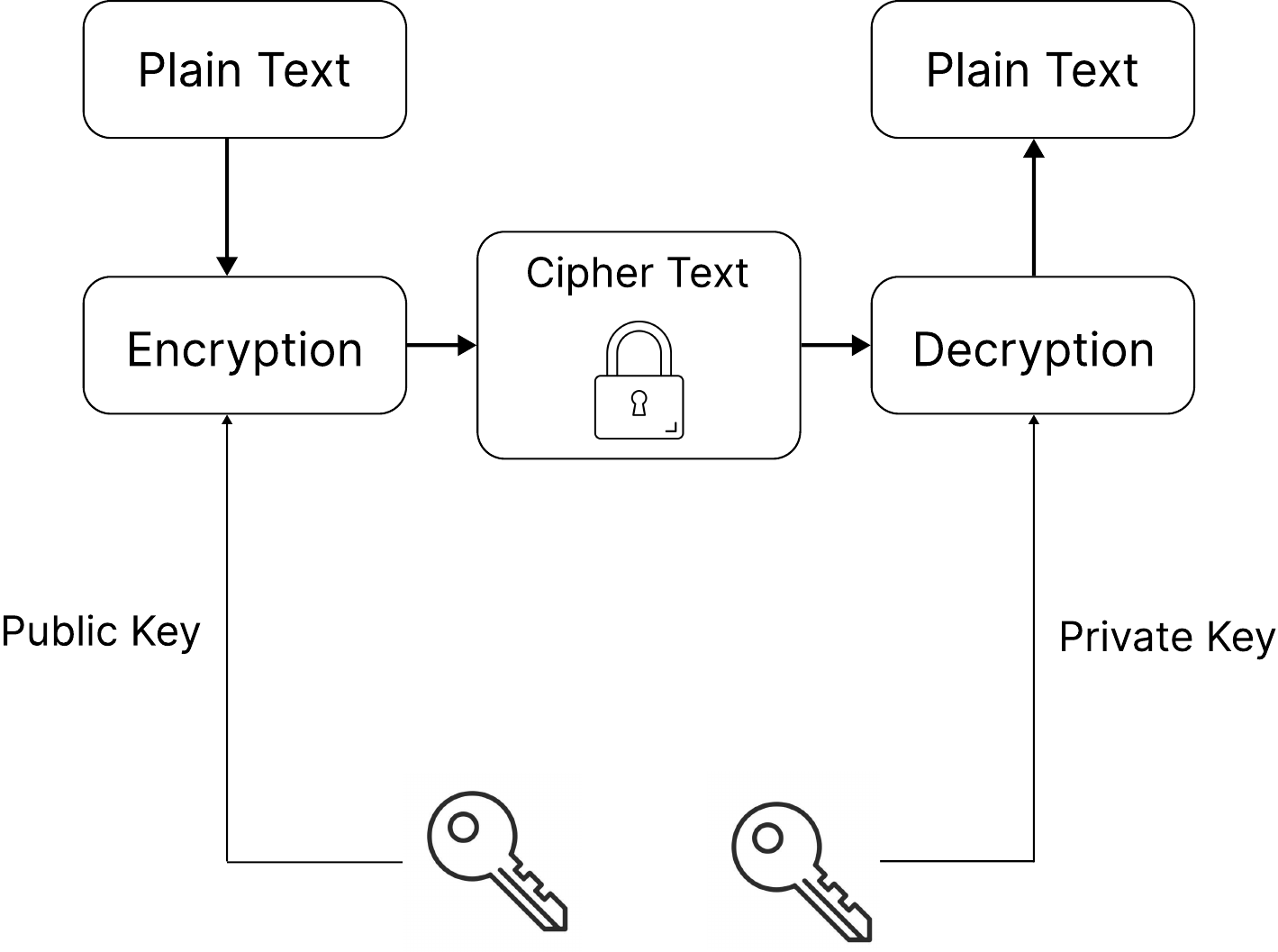}
    \caption{Assymetric Cryptosystem.}
    \label{fig:Assymetric cryptosystem}
\end{figure}

Figure \ref{fig:Assymetric cryptosystem} shows the fundamental workflow of asymmetric cryptography. It illustrates how encryption is performed using a public key, while the corresponding decryption is carried out using the associated private key, ensuring that only the intended recipient can access the original data.
 The plaintext is transformed into ciphertext via the public key, ensuring confidentiality. Because only the entity possessing the corresponding private key can transform the ciphertext back into its original plaintext, this exemplifies the fundamental principle of public‐key encryption in secure communications.

\subsubsection{ElGamal Encryption Scheme}

Based on the discrete logarithm problem's difficulty, the ElGamal encryption scheme ~\cite{elgamal1985public} is a public-key cryptosystem. Operating with a generator \( g \), the scheme functions within a finite cyclic group \( \mathbb{G} \) of prime order \( p \), which forms the mathematical foundation for its security.
 After choosing a private key \( x \in \mathbb{Z}_p^* \), the recipient calculates the matching public key:
\begin{equation}
y = g^x \bmod p
\end{equation}
To encrypt a message \( m \in \mathbb{Z}_p^* \), the sender selects a random ephemeral key \( k \in \mathbb{Z}_p^* \) and computes the ciphertext as:

\begin{equation}
c_1 = g^k \bmod p
\end{equation}
\begin{equation}
c_2 = m \cdot y^k \bmod p
\end{equation}
The ciphertext is the pair \( (c_1, c_2) \). To decrypt, the receiver computes:
\begin{equation}
m = c_2 \cdot (c_1^x)^{-1} \bmod p
\end{equation}
Since \( c_1^x = g^{kx} = y^k \), decryption correctly recovers the plaintext. The use of a random \( k \) makes ElGamal probabilistic, ensuring semantic security under chosen plaintext attacks.

\subsubsection{Least Significant Bit (LSB) steganography }
By gently altering its binary representation without noticeably compromising visual quality, the least significant bit (LSB) steganography technology embeds hidden information into the least significant bits of digital data, such as image pixels. One of the first formal analyses of LSB embedding was given by Fridrich et al.~\cite{fridrich2000lsb}, who modeled the effects of inserting payload data into the LSB plane of color or grayscale images on the cover medium's statistical characteristics. By substituting its least significant bit, the LSB technique embeds a message bit \( b \in \{0,1\} \) into a cover pixel value \( p \):
\begin{equation}
p' = (p \land \sim1) \lor b
\end{equation}
where \( p' \) is the modified pixel, and \( \land \), \( \lor \), and \( \sim \) denote the bitwise AND, OR, and NOT operations, respectively. This minimal change ensures imperceptibility, making LSB steganography a foundational technique in image-based covert communication.

\subsubsection{2D Chaotic System: Hénon Map}

The Hénon map~\cite{henon1976two}, a classical two-dimensional discrete-time dynamical system, is well known for its strong nonlinear characteristics and chaotic behavior under specific parameter values. It is mathematically defined as:
\begin{equation}
\begin{aligned}
x_{n+1} &= 1 - a x_n^2 + y_n, \\
y_{n+1} &= b x_n,
\end{aligned}
\end{equation}
where $a \in [1.0, 1.4]$, $b \in (0, 1)$, and the initial conditions $(x_0, y_0) \in (0, 1]^2$. For appropriate choices of $a$ and $b$—typically $a = 1.4$, $b = 0.3$—the system exhibits chaotic dynamics, including high sensitivity to initial values, ergodicity, and bifurcations~\cite{li2003henon}.

The chaotic sequence $\{x_n, y_n\}$ produced by the map can be normalized to generate pseudo-random values using:
\[
S_n = |x_n| \bmod 1,
\]
which can then be applied in cryptographic contexts for generating substitution boxes (S-boxes), permutation vectors, or masking keys~\cite{wang2018image}.

\begin{figure}[ht]
    \centering
    \includegraphics[width=0.9\linewidth]{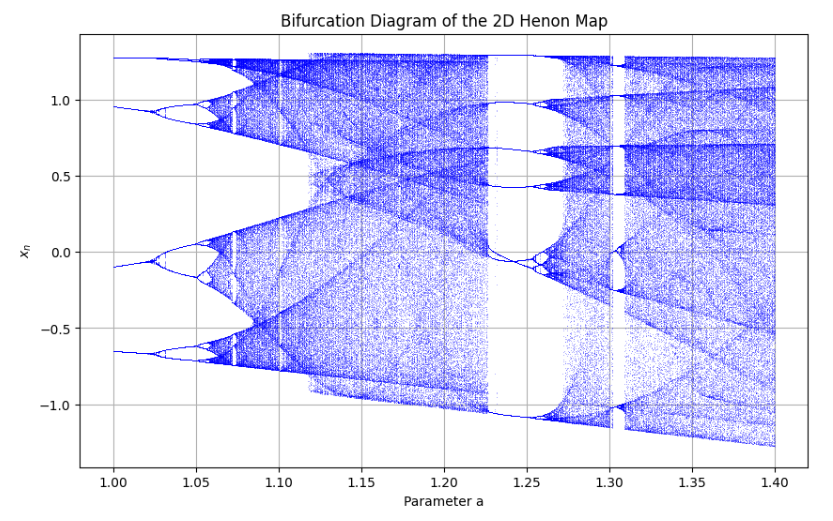}
    \caption{Bifurcation diagram of the 2D Hénon chaotic system.}
    \label{fig:henon_bifurcation}
\end{figure}

As illustrated in Fig.~\ref{fig:henon_bifurcation}, the Hénon map yields a rich and unpredictable spectrum of $x_n$ values in the range $[-1.2, 1.2]$. This diverse behavior, coupled with its two-dimensional structure, allows it to outperform 1D chaotic maps in terms of diffusion strength, making it highly suitable for secure image encryption applications.

\subsubsection{ Classical AES Algorithm}
\begin{figure}[ht]
    \centering
    \includegraphics[width=0.9\linewidth]{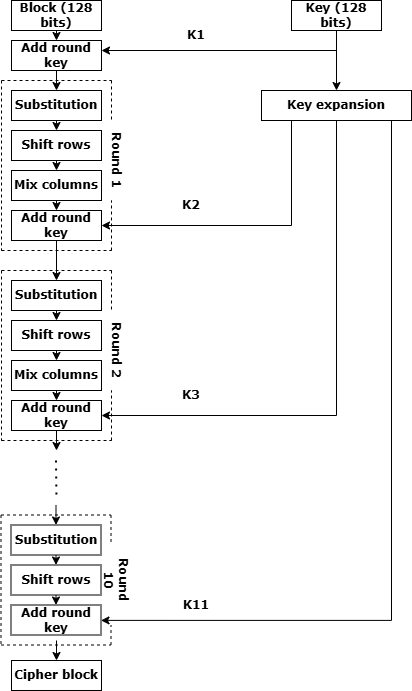}
    \caption{Classic AES algorithm.}
    \label{fig:Classic AES algorithm}
\end{figure}

The Advanced Encryption Standard (AES), formalized by the National Institute of Standards and Technology (NIST) in 2001~\cite{NIST_AES}, is one of the most widely utilized symmetric block ciphers. It operates on 128-bit data blocks and supports key lengths of 128, 192, or 256 bits. The encryption process consists of a series of transformation rounds—10 for 128-bit keys, 12 for 192-bit keys, and 14 for 256-bit keys. Figure~\ref{fig:Classic AES algorithm} illustrates the core cryptographic operations performed during each round.

\vspace{4mm}
\textit{Substitution}\\
\vspace{0.5mm}

The substitution step in AES uses a lookup table known as the S-box to transform each of the 16 input bytes into new values. This process introduces non-linearity into the cipher, making it significantly more resistant to linear and differential cryptanalysis. Each byte in the \(4 \times 4\) state matrix is independently replaced using this fixed 256-entry substitution table~\cite{daemen1999aes}.

\vspace{4mm}
\textit{Shift rows}\\
\vspace{0.5mm}

In the ShiftRows step of AES, the input is treated as a \(4 \times 4\) byte matrix. The first row remains unchanged, while the second, third, and fourth rows are cyclically shifted to the left by one, two, and three positions, respectively. This simple yet effective transformation increases diffusion by rearranging the byte positions across columns, helping to break any predictable structure in the plaintext~\cite{daemen1999aes}.

\vspace{4mm}
\textit{Mix columns}\\
\vspace{0.5mm}

In the MixColumns step of AES, each column of the \(4 \times 4\) state matrix is treated as a polynomial over the finite field \( \text{GF}(2^8) \). This step mixes the data within each column to increase diffusion across the state. Each column is multiplied by a fixed polynomial:
\begin{equation}
a(x) = \{03\}x^3 + \{01\}x^2 + \{01\}x + \{02\}
\end{equation}
modulo \( x^4 + 1 \), where the constants are represented in hexadecimal.

The transformation can be expressed in matrix form as:
\begin{equation}
\begin{bmatrix}
S'_{0,c} \\
S'_{1,c} \\
S'_{2,c} \\
S'_{3,c}
\end{bmatrix}
=
\begin{bmatrix}
\texttt{02} & \texttt{03} & \texttt{01} & \texttt{01} \\
\texttt{01} & \texttt{02} & \texttt{03} & \texttt{01} \\
\texttt{01} & \texttt{01} & \texttt{02} & \texttt{03} \\
\texttt{03} & \texttt{01} & \texttt{01} & \texttt{02}
\end{bmatrix}
\begin{bmatrix}
S_{0,c} \\
S_{1,c} \\
S_{2,c} \\
S_{3,c}
\end{bmatrix}
\end{equation}

Multiplication by \texttt{02} is equivalent to a left shift operation followed by a modular reduction with the primitive polynomial:
\begin{equation}
P(x) = x^8 + x^4 + x^3 + x + 1
\end{equation}
Multiplication by \texttt{01} leaves the byte unchanged, while multiplication by \texttt{03} can be implemented by combining a \texttt{02} multiplication and an XOR with the original byte. These operations help ensure that even small changes in input propagate throughout the state, enhancing the cipher’s resistance to differential attacks~\cite{daemen1999aes}.

\vspace{4mm}
\textit{Add round key}\\
\vspace{0.5mm}

In this step, the round key—generated through the key expansion process—is combined with the state matrix using a bitwise XOR operation, effectively integrating the encryption key into the data~\cite{daemen1999aes}.

\begin{figure}[htbp]
    \centering
    \includegraphics[width=0.9\linewidth]{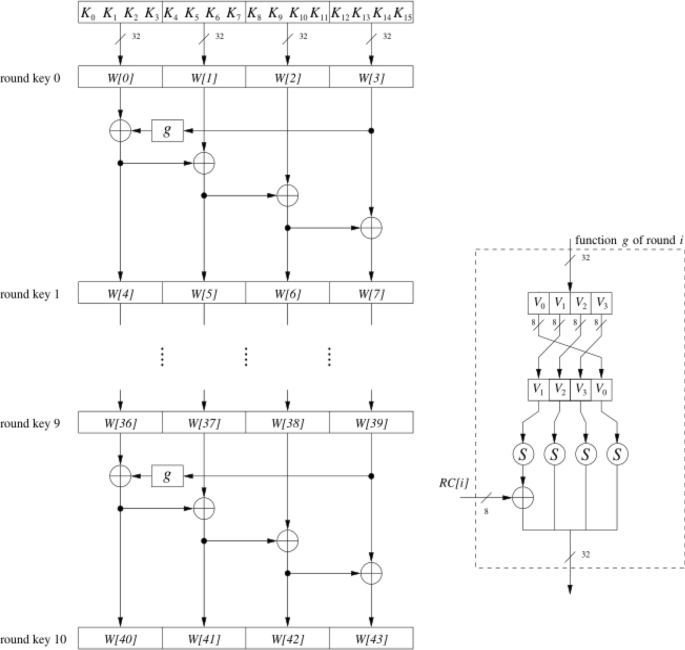}
    \caption{AES key scheduling for keys with a size of 128 bits}
    \label{fig:AES_k}
\end{figure}

\vspace{4mm}
\textit{Key expansion}\\
\vspace{0.5mm}
In the word-oriented AES key schedule, each word represents 32 bits. The expanded key schedule is stored in an array denoted by \( \mathbf{W} \), which consists of several words depending on the original key length. Although the AES key sizes—128, 192, and 256 bits—are structurally similar, each variant follows a slightly different key expansion procedure. Figure~\ref{fig:AES_k} illustrates the key expansion process specifically for 128-bit keys~\cite{hadj2024image}.

The first word of each round key, denoted as \( \mathbf{W}[4i] \), is computed using:
\begin{equation}
\mathbf{W}[4i] = \mathbf{W}[4(i - 1)] + g(\mathbf{W}[4i - 1]),
\end{equation}
where \( g(\cdot) \) is a nonlinear transformation that includes a byte rotation, S-box substitution, and round constant addition. The next three words in the round key are derived as:
\begin{equation}
\mathbf{W}[4i + j] = \mathbf{W}[4i + j - 1] + \mathbf{W}[4(i - 1) + j], \quad j = 1, 2, 3,
\end{equation}
and this process is repeated for \( i = 1, \dots, 10 \) in AES-128.

\subsection{Proposed Image Encryption Scheme }
\begin{figure*}[htbp]
    \centering
    \includegraphics[width=0.8\textwidth, height=.8\textheight]{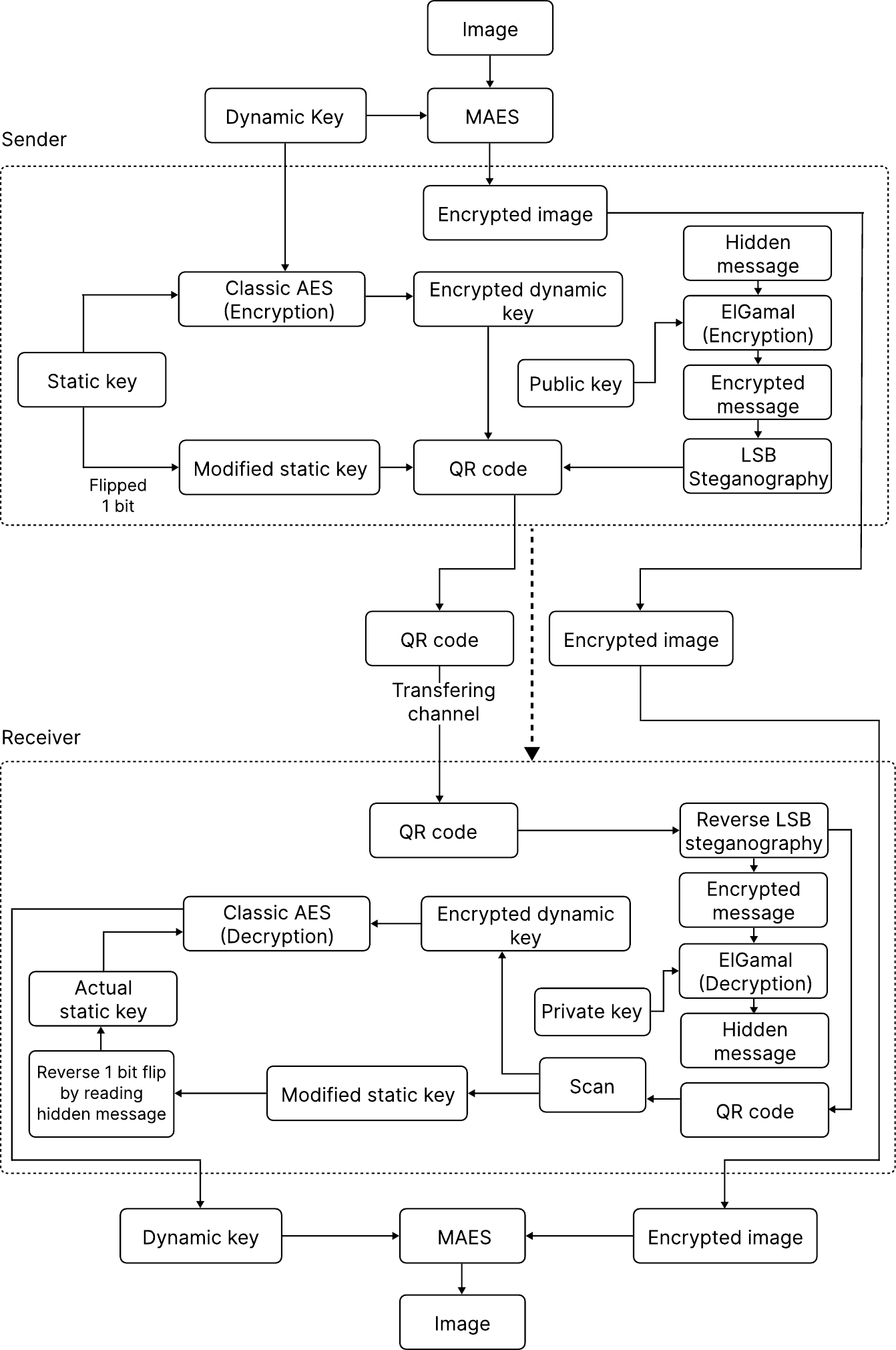}
    \caption{Overall scheme of the proposed framework.}
    \label{fig:aes_key_schedule}
\end{figure*}

There are two main parts to the suggested framework. The first ensures secure key transmission by embedding a subtly modified static key and an encrypted dynamic key within a QR code. This QR code is further encoded with a hidden message and transmitted using a steganographic method to enhance confidentiality and resistance to interception. The second component utilizes the received key in a high-performance, modified AES encryption algorithm for image protection. This enhanced AES incorporates a dynamic S-box derived from a two-dimensional Hénon chaotic map, a shift rows operation guided by a logistic map, and a post-encryption shuffling process to improve diffusion and strengthen resistance against cryptographic attacks. An overview of the complete image encryption framework is presented in Figure~\ref{fig:aes_key_schedule}, highlighting the key components and their interactions.

\begin{figure}[htbp]
    \centering
    \includegraphics[width=0.45\textwidth, height=0.36\textheight]{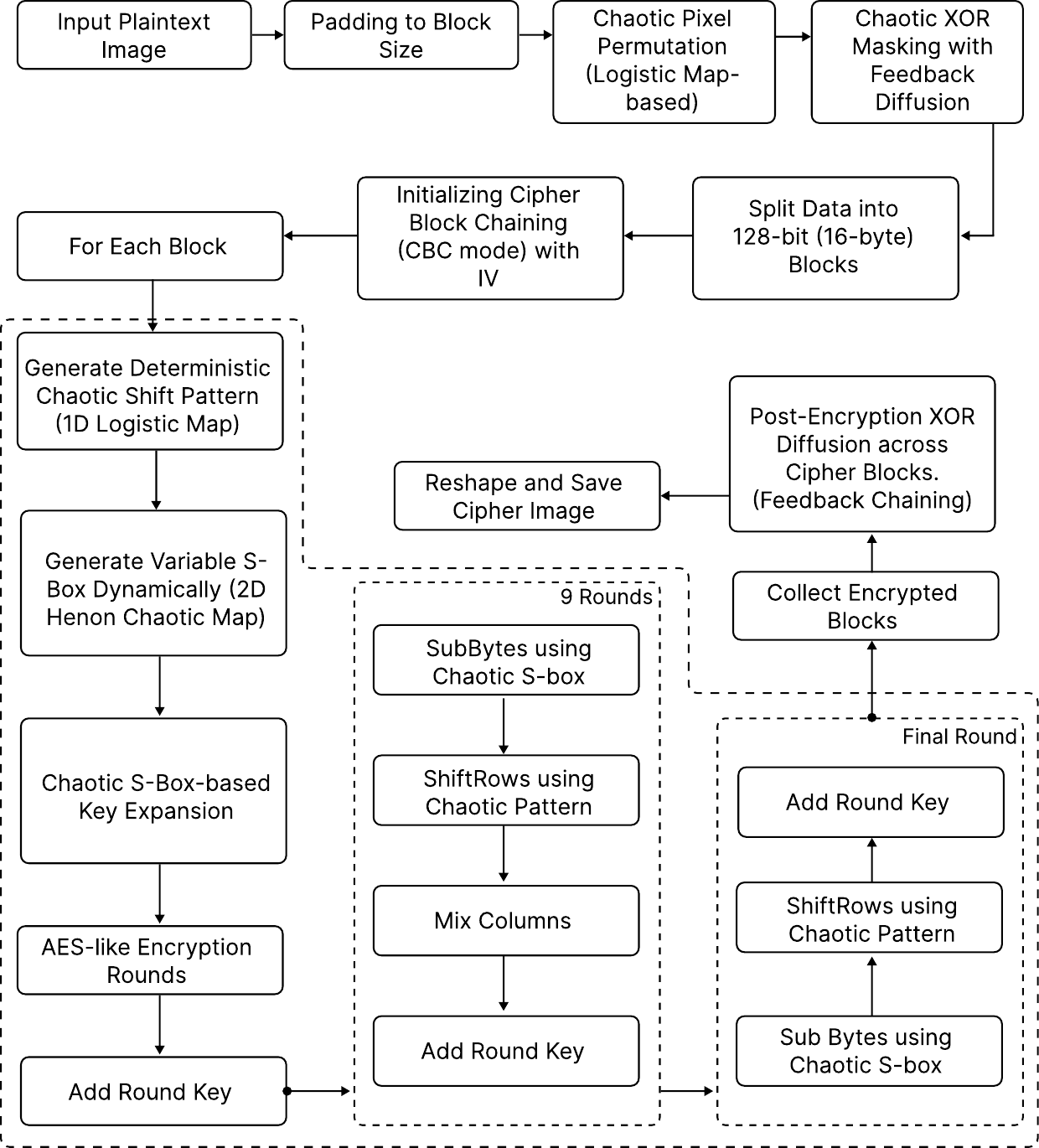}
    \caption{MAES Encryption Pipeline.}
    \label{fig:modified_aes_encr__schedule}
\end{figure}

\begin{figure}[htbp]
    \centering
    \includegraphics[width=0.45\textwidth, height=0.50\textheight]{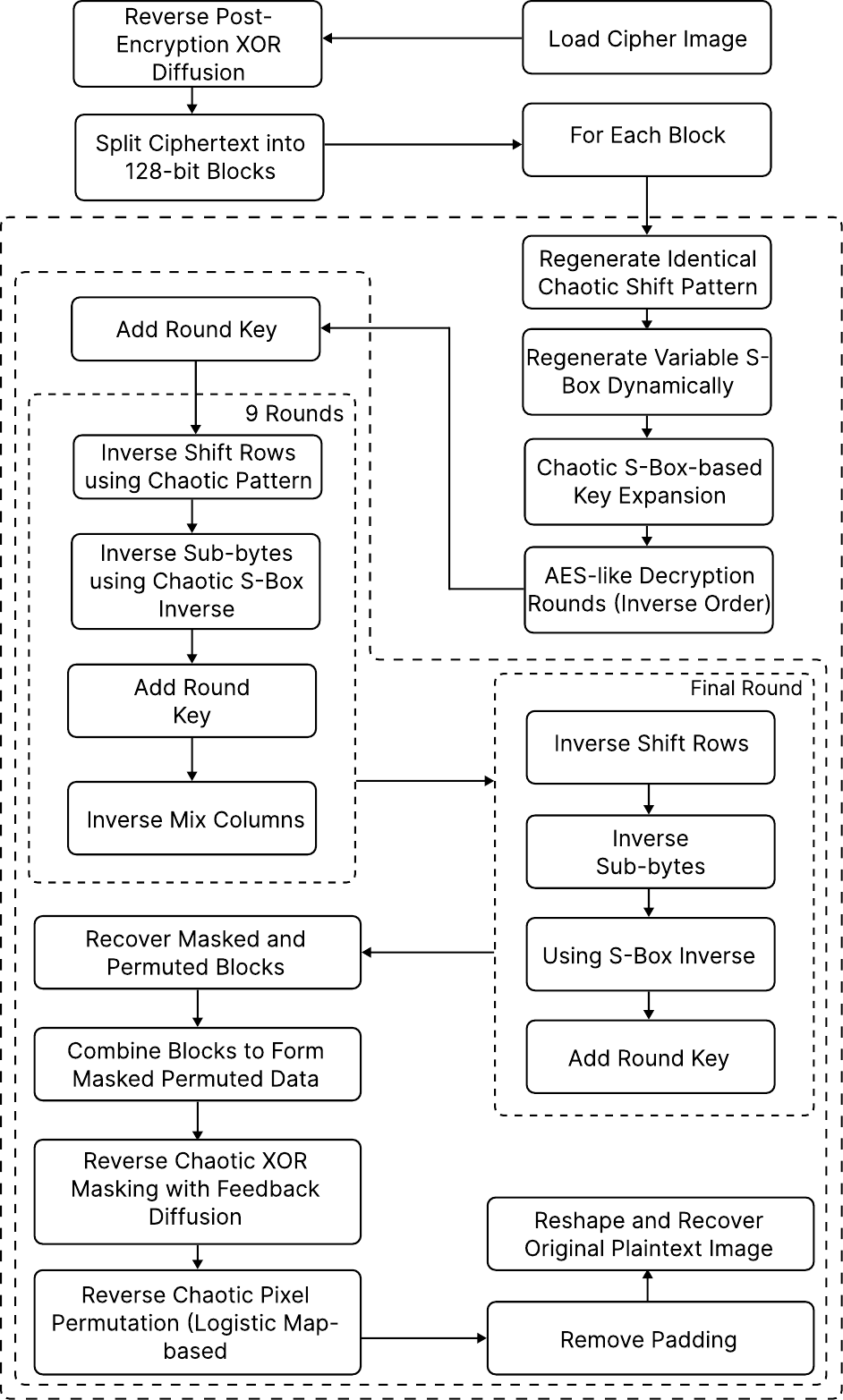}
    \caption{Decryption Pipeline of the Scheme.}
    \label{fig:modified_aes_decp_key_schedule}
\end{figure}

\subsubsection{Key Transmission Methodology in the Proposed Scheme}

The Least Significant Bit (LSB) steganographic approach is used in this methodology to incorporate a concealed message in a QR code without affecting the code's normal readability. In addition to the hidden message, a subtly modified static key and an encrypted dynamic key are encoded into the QR code structure. This approach ensures secure key transmission while preserving the QR code’s scannability and visual integrity. The combined use of steganography and cryptographic encoding forms a dual-layer protection mechanism for robust and covert communication. The complete procedure is outlined in the following steps.

\begin{itemize}
    \item \textbf{Static Key Initialization:} A 128-bit static key is generated and designated for encrypting a dynamic session key (will be used for image encrption using modified AES). Additionally, the static key serves as the foundation for establishing secure communication between the sender and the recipient.

    \item \textbf{Dynamic Key Generation and Encryption:} A randomly generated 128-bit dynamic key is utilized subsequently for image encryption. This key is encrypted in ECB mode using the static key and the traditional AES technique.
    
    \item \textbf{Static Key Bit-Flipping and Trace Encoding:} Prior to embedding, the static key is modified by randomly flipping a single bit, creating a controlled variant. The position of the flipped bit (within the 0–127 range) is recorded.
    
    \item \textbf{Hidden Message Construction:} A short plaintext message is generated such that the sum of its ASCII character values modulo 128 equals the flipped bit position. This modulated message serves as a reversible hint for key recovery.
    
    \item \textbf{ElGamal Encryption of the Modulated Message:} The modulated message is encrypted using the ElGamal encryption scheme and the recipient’s public key, ensuring that only the intended recipient—who possesses the corresponding private key—can decrypt and access the hidden information.

    \item \textbf{QR Code Payload Construction:} A structured JSON payload is constructed that contains the flipped static key (in ASCII), the AES-encrypted dynamic key (in base64), and optionally any metadata. This payload is used to generate a visually scannable QR code.
    
    \item \textbf{Steganographic Embedding of the Hidden Message:} The ElGamal-encrypted modulated message is embedded into the QR code image using Least Significant Bit (LSB) steganography. This technique ensures that the hidden message can be extracted through LSB decoding, while remaining undetectable to standard QR scanners.

    \item \textbf{QR Code Transmission:} The QR code, now carrying both visibly scannable (static and dynamic keys) and hidden (ElGamal-encrypted message) layers, is transmitted to the receiver over a public or semi-trusted channel.
    
    \item \textbf{Receiver-Side QR Decoding and Key Recovery:} Upon receiving the QR code, the receiver extracts the visibly embedded modified static key and the AES-encrypted dynamic key using a standard QR decoder. However, the hidden ElGamal-encrypted message remains undetected at this stage.
    
    \item \textbf{Steganographic Extraction and ElGamal Decryption:} The receiver applies an LSB decoding algorithm on the QR code to extract the hidden encrypted message. Using their private ElGamal key, they decrypt the message to obtain the modulated hint.
    
    \item \textbf{Bit-Flip Position Recovery and Static Key Restoration:} The receiver computes the ASCII sum of the recovered message and applies modulo 128 to retrieve the flipped bit position. By flipping the corresponding bit in the received static key, the original static key is recovered.
    
    \item \textbf{Dynamic Key Decryption and Image Recovery:} The dynamic key that was previously extracted is decrypted using the recovered static key.  Finally, the modified AES encryption algorithm is applied to decrypt the protected image using the recovered dynamic key.

\end{itemize}

This framework presents a secure and covert key transmission mechanism by combining ElGamal encryption, LSB-based steganography, and QR code encoding. By enabling reliable recovery of the original static key through an embedded modulated message, the system ensures robust protection of the encryption process and secure image decryption.

\begin{table*}[htbp, align=\flushleft, width=\textwidth]
\centering
\caption{Examples of Hidden Message and Bit Flipped Position}
\label{tab:permutation_table_example}
\begin{tabular}{>{\raggedright\arraybackslash}p{2.0cm}
                >{\raggedright\arraybackslash}p{5cm}
                >{\raggedright\arraybackslash}p{4.0cm}
                >{\raggedright\arraybackslash}p{1.8cm}
                >{\raggedright\arraybackslash}p{2.5cm}|}
\hline
\textbf{Secret Message} & \textbf{ASCII Values} & \textbf{Calculation of Finding Flipped Position} & \textbf{Flipped Position}  \\
\hline
HI & H=72, I=73 & (72+73) \% 128 & 17  \\
HEY & H=72, E=69, Y=89 & (72+69+89) \% 128 & 102  \\
GO & G=71, O=79 & (71+79) \% 128 & 22  \\
SHE & S=83, H=72, E=69 & (83+72+69) \% 128 & 96  \\
CAT & C=67, A=65, T=84 & (67+65+84) \% 128 & 88  \\
HELLO & H=72, E=69, L=76, L=76, O=79 & (72+69+76+76+79) \% 128 & 116  \\
GOOD & G=71, O=79, O=79, D=68 & (71+79+79+68) \% 128 & 41  \\
\hline
\end{tabular}
\end{table*}

\subsection{Modified AES in the Proposed Scheme}
The proposed methodology enhances the classical AES-128 encryption algorithm by integrating chaotic systems to improve confusion, diffusion, and performance. A key modification involves the use of dynamic S-boxes generated from the 2D Henon map \cite{2d_henon_map}, replacing the fixed AES S-box to introduce key-dependent variability in the SubBytes and key expansion steps. Additionally, the standard ShiftRows operation is replaced with a dynamic shift pattern derived from a logistic map, varying per block to further increase complexity. To strengthen diffusion, chaotic pre- and post-processing steps are applied: plaintext is scrambled using chaotic pixel permutation and XOR diffusion before encryption, and ciphertext blocks undergo an additional XOR chaining process after encryption. With an initialization vector (IV) and a post-cipher XOR diffusion that further distributes block dependencies, the cipher functions in Cipher Block Chaining (CBC) mode. To maintain high efficiency despite the added complexity, the implementation leverages just-in-time (JIT) compilation via \texttt{Numba} for the generation of chaotic sequences. This optimized approach achieves enhanced security features while maintaining performance levels comparable to standard AES implementations.

The next section provides a comprehensive breakdown of each component in the proposed methodology, detailing the algorithmic steps, their underlying rationale, and how they collectively contribute to improved security and performance.
 Key equations describing the chaotic transformations and AES modifications are included, along with pseudocode illustrating the overall encryption process. Algorithm \ref{alg:proposed_encryption_scheme} presents a pseudo-code of the proposed modified AES.

\begin{algorithm}[htbp]
\caption{Variable S-box Generation from 2D Hénon Map}
\label{alg:variable_S_box}
\SetAlgoLined
\KwIn{$x_0$, $y_0$, \texttt{size} (default = 256)}
\KwOut{$SBox$ of length \texttt{size}}

Set Hénon parameters $a \gets 1.4$, $b \gets 0.3$ \\
Initialize empty \texttt{sequence} \\
\For{$i \gets 1$ \KwTo $2 \times \texttt{size}$}{
    $x_{\text{new}} \gets 1 - a \cdot x_0^2 + y_0$ \\
    $y_{\text{new}} \gets b \cdot x_0$ \\
    Append fractional part of $x_{\text{new}}$ to \texttt{sequence} \\
    $x_0 \gets x_{\text{new}}, \quad y_0 \gets y_{\text{new}}$ \\
}
Normalize \texttt{sequence} to range $[0, \texttt{size}-1]$ \\
Remove duplicates, preserving order \\
$SBox \gets$ first \texttt{size} unique values from \texttt{sequence} \\
\KwRet{$SBox$}
\end{algorithm}

\vspace{1em}

\begin{algorithm}[htbp]
\caption{Deterministic Shift Pattern Generation Using Logistic Map}
\label{alg:shift_pattern}
\SetAlgoLined
\KwIn{\texttt{seed}, $r$, \texttt{block\_index}}
\KwOut{\texttt{pattern} of 4 shift values}

$x \gets (\texttt{seed} + \texttt{block\_index} \cdot \texttt{small\_constant}) \bmod 1$ \\
Initialize empty \texttt{pattern} \\
\For{$i \gets 1$ \KwTo $4$}{
    $x \gets r \cdot x \cdot (1 - x)$ \\
    $\texttt{shift\_value} \gets \lfloor x \cdot 4 \rfloor \bmod 4$ \\
    \If{$\texttt{shift\_value} \notin \texttt{pattern}$}{
        Append $\texttt{shift\_value}$ to \texttt{pattern} \\
    }
}
\If{$\texttt{length(pattern)} < 4$}{
    Fill with remaining values from $\{0, 1, 2, 3\} \setminus \texttt{pattern}$
}
\KwRet{\texttt{pattern}}
\end{algorithm}

\subsubsection{Chaotic Sequence Generators}
Chaotic sequence generators—such as the Logistic Map, Hénon Map, and the hybrid Logistic-Tent Map—are employed to introduce deterministic randomness into the encryption process, thereby enhancing security through improved confusion and diffusion. These chaotic systems are favored for their sensitivity to initial conditions, ergodicity, and ability to produce complex and unpredictable sequences~\cite{chen2004new, pareek2006image}. The following sections describe the specific role of each map and justify their integration into the proposed encryption framework.

\begin{algorithm}[htbp]
\caption{Proposed Image Encryption and Decryption Scheme}
\label{alg:proposed_encryption_scheme}
\SetAlgoLined
\SetKwInOut{Input}{Input}
\SetKwInOut{Output}{Output}

\Input{Plaintext image, AES key}
\Output{Encrypted image / Decrypted image}

\BlankLine
\textbf{-- Encryption --} \\
Convert image to grayscale and pad to 16-byte boundary \\
Apply chaotic pixel permutation (logistic map) \\
Apply chaotic XOR mask with feedback (logistic map) \\
Initialize CBC mode with IV \\

\ForEach{16-byte block}{
    Generate variable S-box (2D Hénon map) \\
    CBC pre-XOR with previous ciphertext block \\
    Generate shift pattern (logistic map) \\
    AES encrypt using dynamic S-box and shift pattern \\
}

Apply post-encryption XOR chaining across ciphertext blocks \\
Output: Encrypted image \\

\BlankLine
\textbf{-- Decryption --} \\
Load encrypted image and undo post-encryption XOR chaining \\

\ForEach{ciphertext block}{
    Generate variable S-box and shift pattern \\
    AES decrypt using dynamic S-box and shift pattern \\
    CBC post-XOR with previous ciphertext block \\
}

Inverse chaotic XOR mask and pixel permutation \\
Output: Decrypted image

\end{algorithm}

\subsubsection*{Logistic Map (1D)}

The Logistic Map is employed to generate pseudorandom sequences that drive key operations such as pixel permutation, XOR-based diffusion, and dynamic ShiftRows transformations. It is mathematically defined as:
\begin{equation}
x_{n+1} = r \cdot x_n \cdot (1 - x_n), \quad \text{with } r = 3.99
\end{equation}
To ensure diversity across encryption blocks, the initial seed is slightly perturbed for each block:
\begin{equation}
x_0^{(i)} = (\text{seed} + 0.0001 \cdot i) \bmod 1
\end{equation}
The resulting values are scaled to the byte range as:
\begin{equation}
k_i = \lfloor 256 \cdot x_i \rfloor
\end{equation}
These generated sequences are used to determine the indices for permutation, construct keystreams for XOR diffusion, and dynamically configure ShiftRows operations~\cite{pareek2006image}.

\subsubsection*{Hénon Map (2D)}

The Hénon map dynamically generates the AES substitution box (S-box) using the following equations:
\begin{align}
x_{n+1} &= 1 - a x_n^2 + y_n, \label{eq:henon1} \\
y_{n+1} &= b x_n, \quad a=1.4, \quad b=0.3 \label{eq:henon2}
\end{align}
As described in Equations~\eqref{eq:henon1} and~\eqref{eq:henon2}, the map evolves based on nonlinear recurrence~\cite{chen2004new, wu2018image}. Fractional parts produce a sorted permutation \( \pi \):
\begin{equation}
s_n = |x_n| \bmod 1, \quad S(b) = \pi_b \label{eq:henon3}
\end{equation}
As shown in Equation~\eqref{eq:henon3}, this allows for a dynamic S-box generation process. Chaotic seeds update deterministically per block, ensuring each S-box remains unique and invertible.

\subsubsection*{Logistic-Tent Hybrid Map}

The Logistic–Tent hybrid map accelerates chaotic sequence generation:
\begin{equation}
x_{n+1} =
\begin{cases}
4x_n(1 - x_n), & \text{if } x_n < 0.5 \\
\mu(1 - |2x_n - 1|), & \text{if } x_n \geq 0.5
\end{cases}
\end{equation}
Optimized for fast execution, this hybrid map combines the behaviors of the logistic and tent maps, making it well-suited for high-performance cryptographic applications~\cite{hua2019novel, song2017novel}.

The integration of 1D and 2D chaotic maps enables the generation of dynamic, deterministic, and key-sensitive sequences essential for secure encryption. Each map contributes to specific cryptographic functions—permutation, substitution, and diffusion—ensuring variability across blocks. This multi-map strategy enhances overall complexity while maintaining efficient implementation.

\subsubsection{Chaotic Preprocessing}
Prior to encryption, the plaintext (such as an image array) is subjected to two preprocessing steps that utilize logistic-map-based chaos to effectively obscure underlying patterns.

\subsubsection*{Chaotic Pixel Permutation}

The plaintext bytes undergo permutation guided by logistic map-generated sequences. Iterating the logistic map  times produces a chaotic sequence , sorted to form a permutation  such that:
\begin{equation}
x_{\pi(1)} \leq x_{\pi(2)} \leq \cdots \leq x_{\pi(N)}
\end{equation}
Each byte at position  is relocated to position , introducing key-dependent byte shuffling and ensuring lossless decryption via stored permutation indices.

\subsubsection*{Chaotic XOR Mask with Feedback}

The permuted plaintext is XOR-masked using a keystream generated by the logistic map, incorporating a feedback diffusion mechanism:
\begin{align}
M[i] &= 
\begin{cases}
P[0] \oplus K[0], & i = 0 \\
P[i] \oplus K[i] \oplus M[i-1], & 1 \leq i \leq N-1
\end{cases}
\end{align}
This feedback structure significantly enhances diffusion, ensuring that even a single-bit change in the plaintext propagates throughout the ciphertext. The decryption process applies the inverse operations accordingly. Such feedback-based diffusion schemes are widely used in chaos-based encryption to improve resistance against differential attacks~\cite{wu2011image}.

The proposed preprocessing steps---logistic-map-based permutation and feedback XOR diffusion---transform the plaintext into a highly randomized, noise-like form prior to AES encryption. By disrupting spatial and statistical patterns before the first encryption round, these methods ensure that the data entering the AES core lacks any recognizable structure. This pre-encryption chaos significantly raises the entropy of both the intermediate and final ciphertext outputs. Moreover, because the chaotic sequences are key- and session-dependent, encrypting identical plaintexts with different seeds results in entirely different intermediate states and ciphertexts. This layered approach to confusion and diffusion strengthens the cipher’s resistance to cryptanalytic attacks and enhances overall security.

\subsubsection{Dynamic ShiftRows Pattern via Logistic Map}

To implement a dynamic \textit{ShiftRows} operation, the logistic map is utilized to generate a unique shift pattern for each encryption block. The logistic map iterations are defined as:
\begin{equation}
x_{n+1} = r \cdot x_n \cdot (1 - x_n)
\end{equation}
with initialization parameters:
\begin{equation}
r = 3.99, \quad x_0^{(i)} = (\text{seed} + 0.0001 \cdot i) \bmod 1
\end{equation}
For each encryption block, four successive iterations produce cyclic shift values as follows:
\begin{equation}
\text{shift\_pattern}_r = \lfloor 4 \cdot x_r \rfloor \bmod 4, \quad r = 0, 1, 2, 3
\end{equation}
Each row of the AES state matrix is cyclically shifted based on the corresponding value in the generated shift pattern, enabling a dynamic and block-specific transformation. This enhances diffusion and thwarts pattern-based attacks. The pseudo-code for this process is presented in Algorithm~\ref{alg:shift_pattern}. Such adaptive ShiftRows mechanisms have been shown to improve security in chaos-enhanced AES variants~\cite{hua2019novel}.

\subsubsection{Dynamic S-Box Generation via Henon Map}

The dynamic S-box generation utilizes the chaotic Hénon map to produce a substitution box that varies with each encryption block. This block dependency enhances security by preventing the reuse of a fixed substitution pattern. The Hénon map is defined by the following iterative equations:
\begin{align}
x_{n+1} &= 1 - a x_n^2 + y_n, \\
y_{n+1} &= b x_n
\end{align}
where typical parameters are \( a = 1.4 \) and \( b = 0.3 \). Due to its strong sensitivity to initial conditions and nonlinear behavior, the Hénon map is well-suited for cryptographic applications such as S-box construction~\cite{chen2004new, wu2018image}.

with parameters:
\begin{equation}
a=1.4,\quad b=0.3, \quad x\_0=0.1, \quad y\_0=0.1
\end{equation}
The fractional parts of the generated sequence $\{x_n\}$ are extracted and sorted to form a permutation $\pi$, mapping each byte value to a unique substituted value:
\begin{equation}
S(b) = \pi(b), \quad \forall b \in \mathbb{F}_{2^8}
\end{equation}

After generating each S-box, the chaotic map seeds $x_0, y_0$ are deterministically updated per block to maintain the uniqueness and unpredictability of substitution boxes throughout the encryption process. Algorithm \ref{alg:variable_S_box} presents a pseudo-code of discription.

\subsubsection{Modified AES}
The proposed algorithm's modified AES employs logistic map-based permutations to include new diffusion mechanisms and dynamically created chaotic S-boxes formed from a two-dimensional Hénon map in place of the traditional S-box.  The following are how the modified AES functions:

\begin{itemize}
    \item Instead of using the fixed pattern used in traditional AES, the shift rows operation employs a dynamic shift pattern produced by a logistic map.
    \item The mix columns algorithm functions similarly to the traditional AES algorithm.
    \item Ten sub-keys are created for the encryption rounds by the key expansion operation using a chaotic S-box that is created from a 2D Hénon map.
    \item The substitution operation replaces each byte with a corresponding value from the dynamic S-box, which changes per block or round.
    \item The current state and the round key obtained from the chaotic key schedule are subjected to XOR in the add round key operation.
    \item The post-encryption diffusion operation performs an additional XOR between each ciphertext block and its previous block to strengthen ciphertext randomness.
    \item The feedback chaining mechanism introduces inter-block dependency by propagating changes from one ciphertext block to the next, increasing resistance to differential and statistical attacks.
\end{itemize}

The integration of multiple chaotic maps—including the Logistic, Hénon, and Logistic-Tent maps—facilitates dynamic S-box generation, key-dependent pixel permutation, and feedback-based diffusion, thereby significantly enhancing the cryptographic strength of the AES framework. These chaotic components introduce high entropy, exhibit strong sensitivity to initial conditions, and substantially improve resistance to statistical and differential attacks. Both Figure~\ref{fig:modified_aes_encr__schedule} and Figure~\ref{fig:modified_aes_decp_key_schedule} show the full encryption and decryption pipelines putting these mechanisms into practice.

\section{ Results and Analysis}

\begin{table*}[htbp]
\centering
\caption{Correlation Comparison (Before and After Pixel Shuffling)}
\label{tab:correlation-2d-shifting}
\begin{tabular}{
  >{\raggedright\arraybackslash}p{2.5cm} 
  >{\centering\arraybackslash}p{1.8cm} 
  >{\centering\arraybackslash}p{1.8cm} 
  >{\centering\arraybackslash}p{1.8cm} 
  >{\centering\arraybackslash}p{1.8cm} 
  >{\centering\arraybackslash}p{1.8cm} 
  >{\centering\arraybackslash}p{1.8cm}
}
\toprule
\textbf{Images} & \multicolumn{3}{c}{\textbf{Before Shuffling}} & \multicolumn{3}{c}{\textbf{After Shuffling}} \\
\cmidrule(lr){2-4} \cmidrule(lr){5-7}
& \textbf{Horizontal} & \textbf{Vertical} & \textbf{Diagonal} 
& \textbf{Horizontal} & \textbf{Vertical} & \textbf{Diagonal} \\
\midrule
Lena         & -0.0001 &  0.0040 & -0.0009 & -0.0004 &  0.0040 &  0.0002 \\
Papers       &  0.0020 &  0.0027 &  0.0010 &  0.0022 &  0.0029 & -0.0011 \\
Brain        & -0.0068 & -0.0085 & -0.0007 &  0.0034 & -0.0027 &  0.0048 \\
FingerPrints &  0.0059 &  0.0021 &  0.0004 &  0.0045 &  0.0059 & -0.0019 \\
Lung         & -0.0036 & -0.0021 &  0.0019 &  0.0022 &  0.0053 &  0.0104 \\
Survalance   &  0.0002 &  0.0045 &  0.0018 &  0.0073 & -0.0015 & -0.0036 \\
Monalisa     & -0.0025 &  0.0062 & -0.0026 & -0.0013 & -0.0049 &  0.0110 \\
Plane        & -0.0011 & -0.0101 & -0.0053 &  0.0028 & -0.0033 &  0.0002 \\
Cameraman       &  0.0018 &  0.0009 & -0.0017 & -0.0029 &  0.0003 & -0.0056 \\
Baboon       &  0.0021 & -0.0061 & -0.0064 & -0.0008 &  0.0023 & -0.0058 \\
\bottomrule
\end{tabular}
\end{table*}

\begin{table}[htbp]
\centering
\caption{Security Analysis: Original vs. Encrypted Image}
\label{tab:security_analysis}
\begin{tabular}{lccc}
\toprule
\textbf{Image} & \textbf{Entropy (bits)} & \textbf{NPCR (\%)} & \textbf{UACI (\%)} \\
\midrule
Lena          & 7.9976 & 99.62 & 28.89 \\
Peppers       & 7.9973 & 99.61 & 43.06 \\
Brain         & 7.9973 & 99.60 & 33.56 \\
FingerPrints  & 7.9975 & 99.61 & 46.32 \\
Lung          & 7.9972 & 99.59 & 29.74 \\
Surveillance  & 7.9975 & 99.63 & 28.33 \\
Monalisa      & 7.9968 & 99.62 & 36.41 \\
Plane         & 7.9976 & 99.62 & 30.57 \\
Cameraman        & 7.9971 & 99.62 & 32.04 \\
Baboon        & 7.9968 & 99.65 & 27.33 \\
\bottomrule
\end{tabular}
\end{table}

\begin{table}[htbp]
\centering
\caption{SSIM with 1-Byte Changed AES Key (Key Sensitivity Test)}
\label{tab:key_sensitivity}
\begin{tabular}{lc}
\toprule
\textbf{Image} & \textbf{SSIM} \\
\midrule
Lena          & 0.01105 \\
Peppers       & 0.00856 \\
Brain         & 0.01008 \\
FingerPrints  & 0.00278 \\
Lung          & 0.00870 \\
Surveillance  & 0.00697 \\
Monalisa      & 0.00650 \\
Plane         & 0.01065 \\
Cameraman     & 0.00904 \\
Baboon        & 0.01054 \\
\bottomrule
\end{tabular}
\end{table}

\begin{table}[htbp]
\centering
\caption{NPCR and UACI Results under Differential Attack}
\label{tab:differential_attack}
\begin{tabular}{lcc}
\toprule
\textbf{Image} & \textbf{NPCR (\%)} & \textbf{UACI (\%)} \\
\midrule
Lena          & 99.62 & 49.91 \\
Peppers       & 99.62 & 49.94 \\
Brain         & 99.59 & 49.87 \\
FingerPrints  & 99.60 & 49.96 \\
Lung          & 99.07 & 49.73 \\
Surveillance  & 99.62 & 50.01 \\
Monalisa      & 99.67 & 50.08 \\
Plane         & 99.64 & 50.14 \\
Cameraman     & 99.62 & 49.91 \\
Baboon        & 99.58 & 49.98 \\
\bottomrule
\end{tabular}
\end{table}

\begin{table}[htbp]
\centering
\caption{SSIM after Salt-and-Pepper Noise Attack (Plain vs. Decrypted Noise Image)}
\label{tab:salt_pepper_noise}
\begin{tabular}{lc}
\toprule
\textbf{Image} & \textbf{SSIM} \\
\midrule
Lena          & 0.171394 \\
Peppers       & 0.196000 \\
Brain         & 0.087789 \\
FingerPrints  & 0.163718 \\
Lung          & 0.072885 \\
Surveillance  & 0.322000 \\
Monalisa      & 0.099400 \\
Plane         & 0.108100 \\
Cameraman     & 0.163800 \\
Baboon        & 0.229311 \\
\bottomrule
\end{tabular}
\end{table}

\begin{table}[htbp]
\centering
\caption{SSIM and PSNR after Gaussian Noise Attack (Plain vs. Decrypted Noise Image)}
\label{tab:gaussian_noise}
\begin{tabular}{lcc}
\toprule
\textbf{Image} & \textbf{SSIM} & \textbf{PSNR (dB)} \\
\midrule
Lena          & 0.008192 & 9.12  \\
Peppers       & 0.010420 & 8.65  \\
Brain         & 0.004456 & 8.03  \\
FingerPrints  & 0.001034 & 5.27  \\
Lung          & 0.008695 & 8.87  \\
Surveillance  & 0.009900 & 9.34  \\
Monalisa      & 0.005621 & 7.06  \\
Plane         & 0.009164 & 8.51  \\
Cameraman        & 0.007580 & 8.21  \\
Baboon        & 0.008900 & 9.74  \\
\bottomrule
\end{tabular}
\end{table}

\begin{table}[htbp]
\centering
\caption{Average SSIM and PSNR under 20\% Data Loss (10 Iterations)}
\label{tab:data_loss_updated}
\begin{tabular}{lcc}
\toprule
\textbf{Image} & \textbf{SSIM (mean ± SD)} & \textbf{PSNR (dB ± SD)} \\
\midrule
Lena          & 0.1024 ± 0.0013  & 12.21 ± 0.01  \\
Peppers       & 0.1081 ± 0.0049  & 11.59 ± 0.14  \\
Brain         & 0.1359 ± 0.0095  & 11.16 ± 0.24  \\
FingerPrints  & 0.0848 ± 0.0045  & 8.26 ± 0.15   \\
Lung          & 0.0375 ± 0.0019  & 11.89 ± 0.13  \\
Surveillance  & 0.1807 ± 0.0090  & 12.43 ± 0.14  \\
Monalisa      & 0.0499 ± 0.0040  & 10.04 ± 0.20  \\
Plane         & 0.0675 ± 0.0049  & 11.66 ± 0.25  \\
Cameraman        & 0.0925 ± 0.0042  & 11.17 ± 0.17  \\
Baboon        & 0.1113 ± 0.0080  & 12.56 ± 0.16  \\
\bottomrule
\end{tabular}
\end{table}

\begin{table}[htbp]
\centering
\caption{Homogeneity Analysis of Encrypted Images.}
\label{tab:homogeneity-analysis-doc}
\begin{tabular}{p{3cm} p{3cm}}
\toprule
\textbf{Image} & \textbf{Homogeneity} \\
\midrule
Lena        & 0.3901 \\
Peppers        & 0.3887 \\
Baboon      & 0.3903 \\
Brain       & 0.3881 \\
Fingerprint & 0.3908 \\
Lung        & 0.3894 \\
Surveillance & 0.3889 \\
Monalisa    & 0.3885 \\
Plane   & 0.3891 \\
Cameraman   & 0.3898 \\

\bottomrule
\end{tabular}
\end{table}

\begin{table}[htbp]
\centering
\caption{Energy Analysis in Encrypted and Plain Images.}
\label{tab:energy-analysis-doc}
\begin{tabular}{p{2cm} p{2cm} p{2cm}}
\toprule
\textbf{Image} & \textbf{Plain Image} & \textbf{Encrypted Image} \\
\midrule
Lena        & 0.0458 & 0.0039 \\
Peppers        & 0.0414 & 0.0039 \\
Baboon      & 0.0363 & 0.0039 \\
Brain       & 0.0638 & 0.0039 \\
Fingerprint & 0.6305 & 0.0039 \\
Lung        & 0.0677 & 0.0039 \\
Surveillance & 0.0238 & 0.0039 \\
Monalisa    & 0.1025 & 0.0039 \\
Plane    & 0.0532 & 0.0039 \\
Cameraman   & 0.0941 & 0.0039 \\

\bottomrule
\end{tabular}
\end{table}

\begin{table*}[htbp]
\centering
\caption{Correlation Analysis of Different Encryption Schemes.}
\label{tab:Correlation analysis of different encryption schemes}
\begin{tabular}{p{1.5cm} p{1.5cm} p{1.5cm} p{1.5cm} p{1.5cm} p{1.5cm} p{1.5cm} p{1.5cm} p{1.5cm}}
\toprule
\textbf{Plaintext Image} & \textbf{Direction} & \textbf{Hua \& Zhou \cite{hua_zhou}} & \textbf{Zhu et al. \cite{zhu}} & \textbf{Ping et al. \cite{ping}} & \textbf{Diab \cite{diab}} & \textbf{Pak \& Huang \cite{pak}} & \textbf{Ye \& Huang \cite{ye_huang}} & \textbf{Proposed} \\
\midrule

\multirow{3}{*}{Cameraman} 
& Horizontal & 0.0060 & 0.0015 & 0.0018 & 0.0021 & 0.0015 & 0.0032 & $-0.0029$ \\
& Vertical   & 0.0040 & 0.0035 & 0.0038 & 0.0021 & 0.0035 & 0.0032 & 0.0003 \\
& Diagonal   & 0.0060 & 0.0025 & 0.0038 & 0.0021 & 0.0035 & 0.0032 & $-0.0056$ \\

\midrule

\multirow{3}{*}{Lenna} 
& Horizontal & 0.0039 & 0.0031 & $-0.0221$ & $-0.0046$ & 0.00321 & 0.0022 & $-0.0004$ \\
& Vertical   & 0.0015 & $-0.0021$ & $-0.0035$ & 0.0021 & $-0.0035$ & $-0.0020$ & 0.0040 \\
& Diagonal   & 0.0016 & $-0.0035$ & $-0.0023$ & $-0.0033$ & 0.0042 & 0.0031 & 0.0002 \\

\midrule

\multirow{3}{*}{Baboon} 
& Horizontal & 0.0029 & 0.0031 & $-0.0326$ & $-0.0076$ & 0.00167 & 0.0036 & $-0.0008$ \\
& Vertical   & 0.0015 & $-0.0021$ & $-0.0035$ & 0.0021 & $-0.0035$ & $-0.0020$ & 0.0023 \\
& Diagonal   & 0.0016 & $-0.0035$ & $-0.0023$ & $-0.0033$ & 0.0042 & 0.0031 & $-0.0058$ \\

\midrule

\multirow{3}{*}{Aeroplane} 
& Horizontal & 0.0039 & 0.0031 & $-0.0221$ & $-0.0046$ & 0.00321 & 0.0022 & 0.0028 \\
& Vertical   & 0.0036 & $-0.0065$ & $-0.0040$ & 0.0031 & $-0.0041$ & $-0.0031$ & $-0.0033$ \\
& Diagonal   & 0.0019 & $-0.0039$ & $-0.0030$ & $-0.0040$ & 0.0040 & 0.0039 & 0.0002 \\

\bottomrule
\end{tabular}
\label{tab:correlation}
\end{table*}

\begin{table*}[htbp]
\centering
\caption{Entropy Analysis of Encrypted Images using Different Schemes.}
\label{tab:Entropy analysis of encrypted images.}
\begin{tabular}{p{1.5cm} p{1.75cm} p{1.75cm} p{1.75cm} p{1.75cm} p{1.75cm} p{1.75cm} p{1.75cm} p{1.5cm}}
\toprule
\textbf{Plaintext Image} & \textbf{Hua \& Zhou \cite{hua_zhou}} & \textbf{Zhu et al. \cite{zhu}} & \textbf{Ping et al. \cite{ping}} & \textbf{Diab \cite{diab}} & \textbf{Pak \& Huang \cite{pak}} & \textbf{Ye \& Huang \cite{ye_huang}} & \textbf{Proposed} \\
\midrule
Cameraman               & 7.9671 & 7.9753 & 7.9720 & 7.9735 & 7.9881 & 7.9860 & 7.9971 \\
Lenna                   & 7.9965 & 7.9963 & 7.9865 & 7.9981 & 7.9986 & 7.9971 & 7.9976 \\
Baboon                  & 7.9796 & 7.9880 & 7.9983 & 7.9983 & 7.9975 & 7.9986 & 7.9968 \\
Aeroplane               & 7.9899 & 7.9732 & 7.9734 & 7.9878 & 7.9860 & 7.9856 & 7.9976 \\
\bottomrule
\end{tabular}
\label{tab:entropy-analysis}
\end{table*}

\begin{table*}[htbp]
\centering
\caption{NPCR Analysis.}
\label{tab:npcr-analysis}
\begin{tabular}{p{2.5cm} p{1.8cm} p{1.8cm} p{1.8cm} p{2.0cm} p{2.0cm} p{2.0cm} p{2.0cm}}
\toprule
\textbf{Plaintext Images} & 
\textbf{Neamah \cite{7d_chaotic}} & 
\textbf{Hua et al. \cite{hua2019cosine}} & 
\textbf{Wu et al. \cite{wu2018image}} & 
\textbf{Luo et al. \cite{luo2020image}} & 
\textbf{Hosny et al. \cite{hosny2021new}} &
\textbf{Proposed}\\
\midrule
Aeroplane & 0.9961 & 0.9962 & 0.9963 & 0.9953 & 0.9960 & 0.9964\\
Baboon   & 0.9958 & 0.9961 & 0.9959 & 0.9954 & 0.9957 & 0.9958\\

Peppers  & 0.9958 & 0.9960 & 0.9960 & 0.9960 & 0.9942 & 0.9962\\
\bottomrule
\end{tabular}
\end{table*}

\begin{table*}[htbp]
\centering
\caption{UACI Analysis.}
\label{tab:uaci-analysis}
\begin{tabular}{p{1.5cm} p{1.8cm} p{1.8cm} p{1.8cm} p{1.8cm} p{1.8cm} p{1.8cm} p{1.5cm}}
\toprule
\textbf{Plaintext Image} & \textbf{Hua \& Zhou \cite{hua_zhou}} & \textbf{Zhu et al. \cite{zhu}} & \textbf{Ping et al. \cite{ping}} & \textbf{Diab \cite{diab}} & \textbf{Pak \& Huang \cite{pak}} & \textbf{Ye \& Huang \cite{ye_huang}} & \textbf{Proposed} \\
\midrule
Lenna & 33.5123 & 33.3965 & 33.1965 & 33.4036 & 33.4027 & 33.4013 & 49.91 \\
Cameraman & 33.4136 & 33.3021 & 33.3687 & 33.6031 & 33.6001 & 33.6012 & 49.91 \\
Peppers    & 33.4961 & 33.3024 & 33.3964 & 33.4031 & 33.4012 & 33.6033 & 49.98 \\
Aeroplane & 33.4856 & 33.4021 & 33.4954 & 33.4012 & 33.4035 & 33.4011 & 50.14 \\
\bottomrule
\end{tabular}
\end{table*}

\begin{table*}[htbp]
\centering
\caption{Homogeneity Analysis.}
\label{tab:homogeneity-analysis}
\begin{tabular}{p{1.5cm} p{1.8cm} p{1.8cm} p{1.8cm} p{1.8cm} p{1.8cm} p{1.8cm} p{1.5cm}}
\toprule
\textbf{Plaintext Image} & \textbf{Hua \& Zhou \cite{hua_zhou}} & \textbf{Zhu et al. \cite{zhu}} & \textbf{Ping et al. \cite{ping}} & \textbf{Diab \cite{diab}} & \textbf{Pak \& Huang \cite{pak}} & \textbf{Ye \& Huang \cite{ye_huang}} & \textbf{Proposed} \\
\midrule
Cameraman & 0.5036 & 0.5078 & 0.4706 & 0.6098 & 0.4779 & 0.4878 & 0.3898 \\
Lenna     & 0.5033 & 0.5077 & 0.4216 & 0.5099 & 0.4799 & 0.4998 & 0.3901 \\
Baboon    & 0.5099 & 0.5096 & 0.4763 & 0.5096 & 0.4797 & 0.4888 & 0.3903 \\
Aeroplane & 0.4966 & 0.4996 & 0.4768 & 0.5086 & 0.4896 & 0.4961 & 0.3891 \\
\bottomrule
\end{tabular}
\end{table*}

\begin{table*}[htbp]
\centering
\caption{Energy Analysis.}
\label{tab:energy-analysis}
\begin{tabular}{p{1.5cm} p{1.8cm} p{1.8cm} p{1.8cm} p{1.8cm} p{1.8cm} p{1.8cm} p{1.5cm}}
\toprule
\textbf{Plaintext Image} & \textbf{Hua \& Zhou \cite{hua_zhou}} & \textbf{Zhu et al. \cite{zhu}} & \textbf{Ping et al. \cite{ping}} & \textbf{Diab \cite{diab}} & \textbf{Pak \& Huang \cite{pak}} & \textbf{Ye \& Huang \cite{ye_huang}} & \textbf{Proposed} \\
\midrule
Cameraman & 0.0160 & 0.0159 & 0.0169 & 0.0168 & 0.0165 & 0.0165 & 0.0039 \\
Lenna     & 0.0163 & 0.0162 & 0.0160 & 0.0165 & 0.0160 & 0.0161 & 0.0039 \\
Baboon    & 0.0160 & 0.0161 & 0.0162 & 0.0162 & 0.0163 & 0.0162 & 0.0039 \\
Aeroplane & 0.0160 & 0.0159 & 0.0160 & 0.0162 & 0.0164 & 0.0163 & 0.0039 \\
\bottomrule
\end{tabular}
\end{table*}

\begin{table*}[htbp]
\centering
\caption{MSE Analysis. }
\label{tab:mse-analysis}
\begin{tabular}{p{1.5cm} p{1.8cm} p{1.8cm} p{1.8cm} p{1.8cm} p{1.8cm} p{1.8cm} p{1.5cm}}
\toprule
\textbf{Plaintext Image} & \textbf{Hua \& Zhou \cite{hua_zhou}} & \textbf{Zhu et al. \cite{zhu}} & \textbf{Ping et al. \cite{ping}} & \textbf{Diab \cite{diab}} & \textbf{Pak \& Huang \cite{pak}} & \textbf{Ye \& Huang \cite{ye_huang}} & \textbf{Proposed} \\
\midrule
Cameraman & 5.88 & 8.31 & 6.16 & 5.13 & 3.68 & 6.74 & 0 \\
Lenna     & 6.43 & 9.54 & 5.32 & 3.80 & 4.15 & 3.39 & 0 \\
Baboon    & 3.32 & 8.18 & 4.06 & 3.39 & 3.71 & 4.91 & 0 \\
Aeroplane & 7.39 & 6.05 & 9.95 & 8.19 & 3.38 & 4.96 & 0 \\
\bottomrule
\end{tabular}
\end{table*}

\begin{table*}[htbp]
\centering
\caption{PSNR Analysis.}
\label{tab:psnr-analysis}
\begin{tabular}{p{1.5cm} p{1.8cm} p{1.8cm} p{1.8cm} p{1.8cm} p{1.8cm} p{1.8cm} p{1.5cm}}
\toprule
\textbf{Plaintext Image} & \textbf{Hua \& Zhou \cite{hua_zhou}} & \textbf{Zhu et al. \cite{zhu}} & \textbf{Ping et al. \cite{ping}} & \textbf{Diab \cite{diab}} & \textbf{Pak \& Huang \cite{pak}} & \textbf{Ye \& Huang \cite{ye_huang}} & \textbf{Proposed} \\
\midrule
Cameraman & 202.36 & 209.38 & 189.30 & 198.65 & 192.30 & 216.10 & $\infty$ \\
Lenna     & 216.35 & 218.81 & 196.31 & 207.30 & 209.95 & 207.37 & $\infty$ \\
Baboon    & 207.98 & 239.16 & 207.90 & 219.16 & 207.37 & 213.16 & $\infty$ \\
Aeroplane & 228.73 & 216.38 & 215.97 & 205.19 & 215.76 & 207.20 & $\infty$ \\
\bottomrule
\end{tabular}
\end{table*}

\subsection{Simulation Setup and Parameter Configuration}

The simulations created with Python on a 64-bit machine running Microsoft Windows 10 and equipped with an Intel Core i5-7300HQ Processor (6M Cache, 2.50 GHz up to 3.50 GHz), 8 GB DDR4 RAM, 1TB SATA HDD, and NVIDIA GeForce GTX 1050 (4GB) are shown in this part.  The suggested chaotic AES-based encryption algorithm's performance is assessed using normal grayscale images with a $256 \times 256$ pixel size.

The modified AES-128 algorithm employs the following parameters and beginning conditions: initial seeds $x_0 = 0.5$ for XOR masking, $x_0 = 0.75$ for pre-encryption pixel permutation, and $x_0 = 0.7$ for dynamic ShiftRows pattern generation, with logistic map parameter $r = 3.99$. A two-dimensional Hénon chaotic map with parameters $a = 1.4$, $b = 0.3$, and initial seeds $x_0 = 0.1$, $y_0 = 0.1$ is used to dynamically build the substitution box. In order to maintain unpredictability, the Initialization Vector (IV) is usually created at random while the encryption is in Cipher Block Chaining (CBC) mode. However, a fixed 16-byte IV is employed in this investigation to preserve result reproducibility:
\begin{align*}
\text{IV} = (&23, 145, 67, 89, 12, 200, 34, 222, \\
            &57, 104, 18, 73, 94, 161, 205, 19)
\end{align*}

Although the AES key is generally intended to be dynamically generated for each encryption session to enhance security, a static 128-bit AES key is used in this simulation to ensure consistency of experimental results:
\begin{align*}
\text{Key} = (&43, 40, 171, 9, 126, 174, 247, 207, \\
             &21, 210, 21, 79, 22, 166, 136, 60)
\end{align*}

The proposed encryption framework incorporates multiple layers of chaos-enhanced operations, including pre-AES permutation, XOR masking with feedback diffusion, block-wise variable S-box substitution, and dynamic ShiftRows. A post-encryption logistic permutation step further obfuscates spatial information in the ciphertext image.

\begin{figure*}[htbp]
    \centering
    \includegraphics[width=\textwidth, height=0.74\textheight]{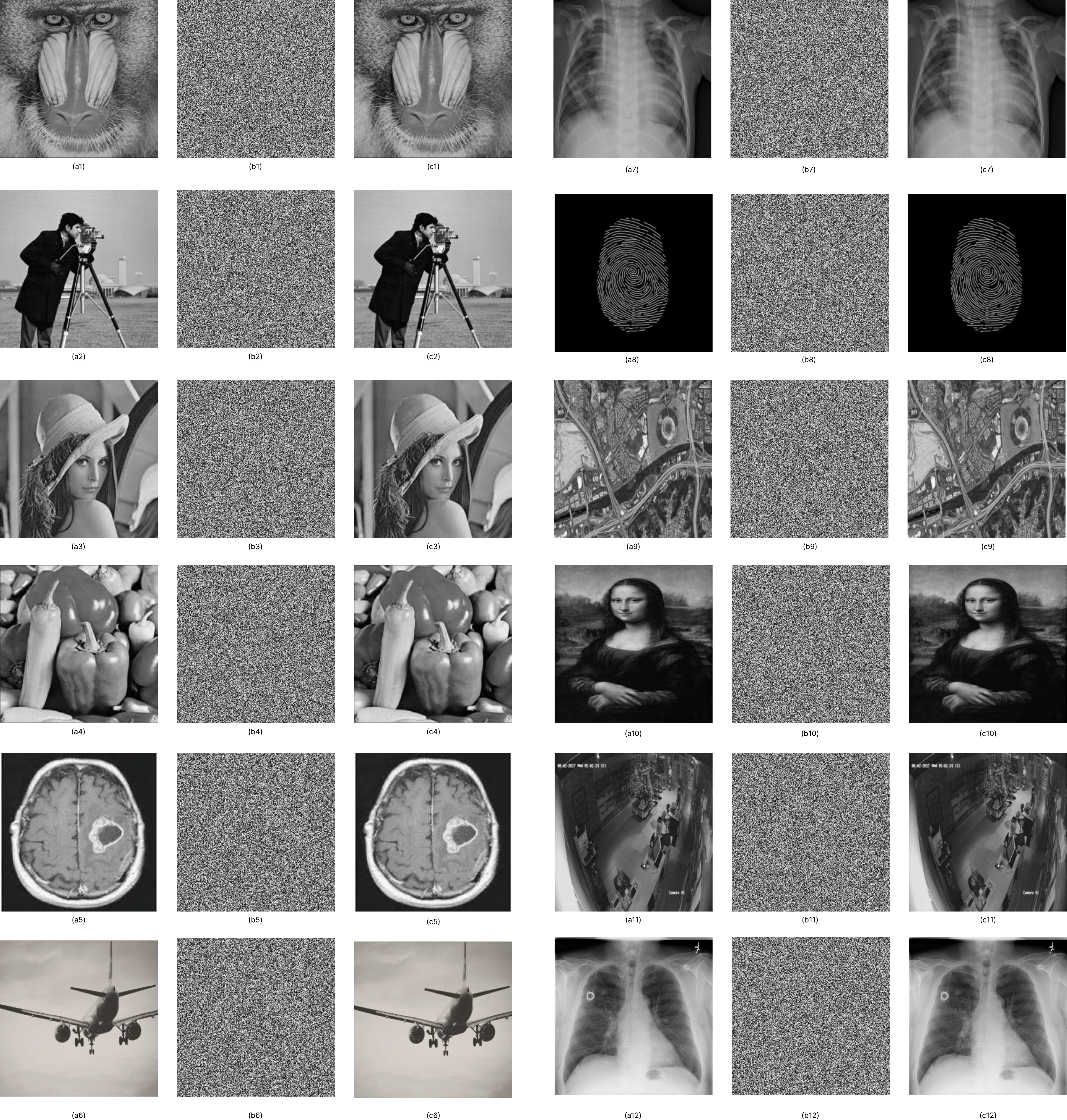}
    \caption{Encryption and Decryption Results: (a\textsubscript{i}) plain images, (b\textsubscript{i}) cipher images, and (c\textsubscript{i}) decrypted images.}
    \label{fig:encdec_grid}
\end{figure*}

The results of encryption and decryption are illustrated in Figure \ref{fig:encdec_grid}. The encrypted outputs appear visually indistinguishable from noise and reveal no discernible structure of the original images. Conversely, the decrypted images perfectly match their respective plaintext versions, confirming the algorithm's effectiveness in maintaining high security while preserving lossless recovery.

\subsection{QR Code Robustness}

To assess the resilience of the proposed QR code-based key distribution method under practical transmission conditions, we conducted a series of distortion experiments involving JPEG compression (at quality levels 90, 70, 50, and 30), Gaussian noise injection ($\sigma = 20$), and Gaussian blur (5$\times$5 kernel). In each case, the visually scannable QR code containing the modified static key and the AES-encrypted dynamic key was successfully decoded without any data loss or corruption. The recovered payload remained identical across all distortions, confirming that the QR code retained full integrity even under significant degradation. This robustness is attributed to the use of error correction level \texttt{H} in QR generation, which enables reliable decoding despite visual noise or compression artifacts. These results validate the practicality of the proposed dual-key transmission scheme in real-world scenarios, such as messaging platforms or low-quality image scans, and address a critical gap in prior works that did not evaluate QR stability under such conditions. A summary of decoding success under each distortion type is presented in \textbf{Table~\ref{tab:qr_robustness}}, which confirms 100\% recovery accuracy across all test cases.

\begin{table}[ht]
\centering
\caption{QR Code Decoding Success Under Various Distortions}
\label{tab:qr_robustness}
\begin{tabular}{l|p{1.5cm}}
\hline
\textbf{Distortion Type} & \textbf{Successfully Decoded} \\
\hline
JPEG Compression (Q = 90) & \ding{51} \\
JPEG Compression (Q = 70) & \ding{51} \\
JPEG Compression (Q = 50) & \ding{51} \\
JPEG Compression (Q = 30) & \ding{51} \\
Gaussian Noise ($\sigma = 20$) & \ding{51} \\
Gaussian Blur ($5 \times 5$) & \ding{51} \\
Salt \& Pepper Noise (Density = 0.05) & \ding{51} \\
\hline
\end{tabular}
\end{table}

\subsection{ Histogram Analysis}
An image's histogram, which represents the picture's structural features, is one of the best ways to depict the distribution of pixel intensities in the image \cite{ris1}. A basic image's histogram usually shows an uneven pattern that is specific to that image. Consequently, a robust encryption algorithm should transform this distribution such that the pixel values in the encrypted image are uniformly spread, minimizing any detectable patterns \cite{ris2}.

\begin{figure*}[htbp]
    \centering
    \includegraphics[width=0.85\textwidth, height=.8\textheight]{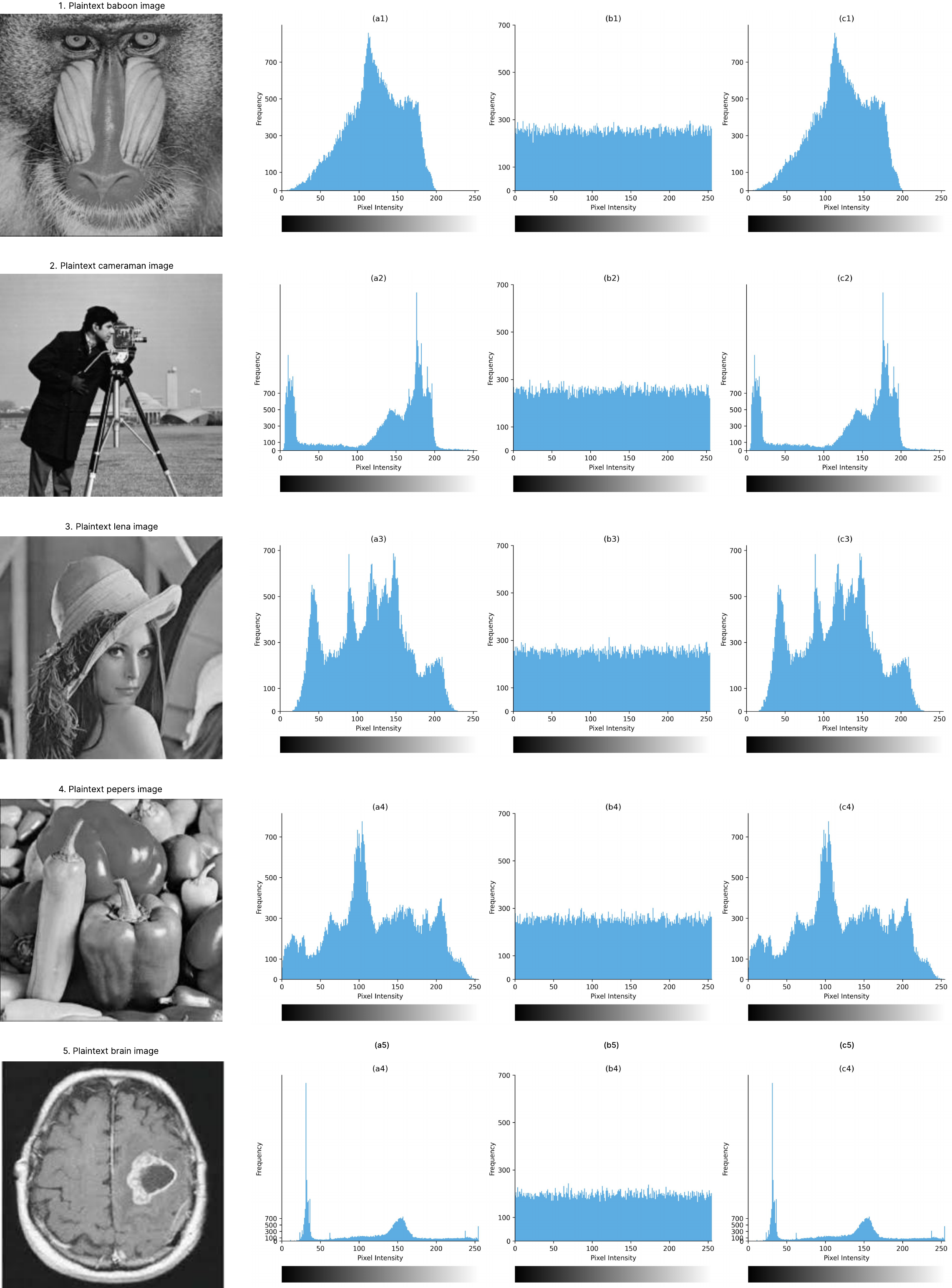}
    \caption{Histogram of: (a\textsubscript{i}) plain images, (b\textsubscript{i}) cipher images, (c\textsubscript{i}) decrypted images}
    \label{fig:histograms}
\end{figure*}

The histograms of the original (plain) photographs, their matching encrypted (cipher) images, and the decrypted images using the suggested modified AES-128 technique are shown in Figure \ref{fig:histograms}. Effective diffusion is indicated by the virtually uniform distribution of the encrypted images' histograms. Additionally, lossless recovery is confirmed by the decrypted images' histograms closely resembling the original plain images' histograms. These findings show that the modified AES-128 algorithm offers robust defense against attacks based on histograms.

\subsection{ Correlation Coefficients Analysis}

In natural images, neighboring pixels typically exhibit strong correlations, which can unintentionally reveal patterns or structures. To ensure confidentiality, an effective image encryption scheme must minimize these correlations as much as possible. The correlation between two adjacent pixels \( x \) and \( y \) is measured using the correlation coefficient \( r_{xy} \), given in Equation~\eqref{eq:corr_coef}~\cite{wu2011image}:

\begin{equation}
r_{xy} = \frac{\text{Cov}(x, y)}{\sqrt{D(x)} \sqrt{D(y)}}
\label{eq:corr_coef}
\end{equation}

Here, the covariance \(\text{Cov}(x, y)\), mean \(E(x)\), and variance \(D(x)\) are computed using the following expressions:

\begin{equation}
\text{Cov}(x, y) = \frac{1}{N} \sum_{i=1}^{N} (x_i - E(x))(y_i - E(y))
\label{eq:cov}
\end{equation}

\begin{equation}
E(x) = \frac{1}{N} \sum_{i=1}^{N} x_i
\label{eq:expectation}
\end{equation}

\begin{equation}
D(x) = \frac{1}{N} \sum_{i=1}^{N} (x_i - E(x))^2
\label{eq:variance}
\end{equation}

The correlation coefficient ranges from \(-1\) to \(1\), where values close to \(1\) or \(-1\) indicate strong positive or negative correlation, respectively, while values near zero suggest weak or no correlation. In encrypted images, low correlation values are desirable, as they imply effective decorrelation and higher security.

\begin{figure*}[htbp]
    \centering
    \includegraphics[width=0.9\textwidth, height=0.8\textheight]{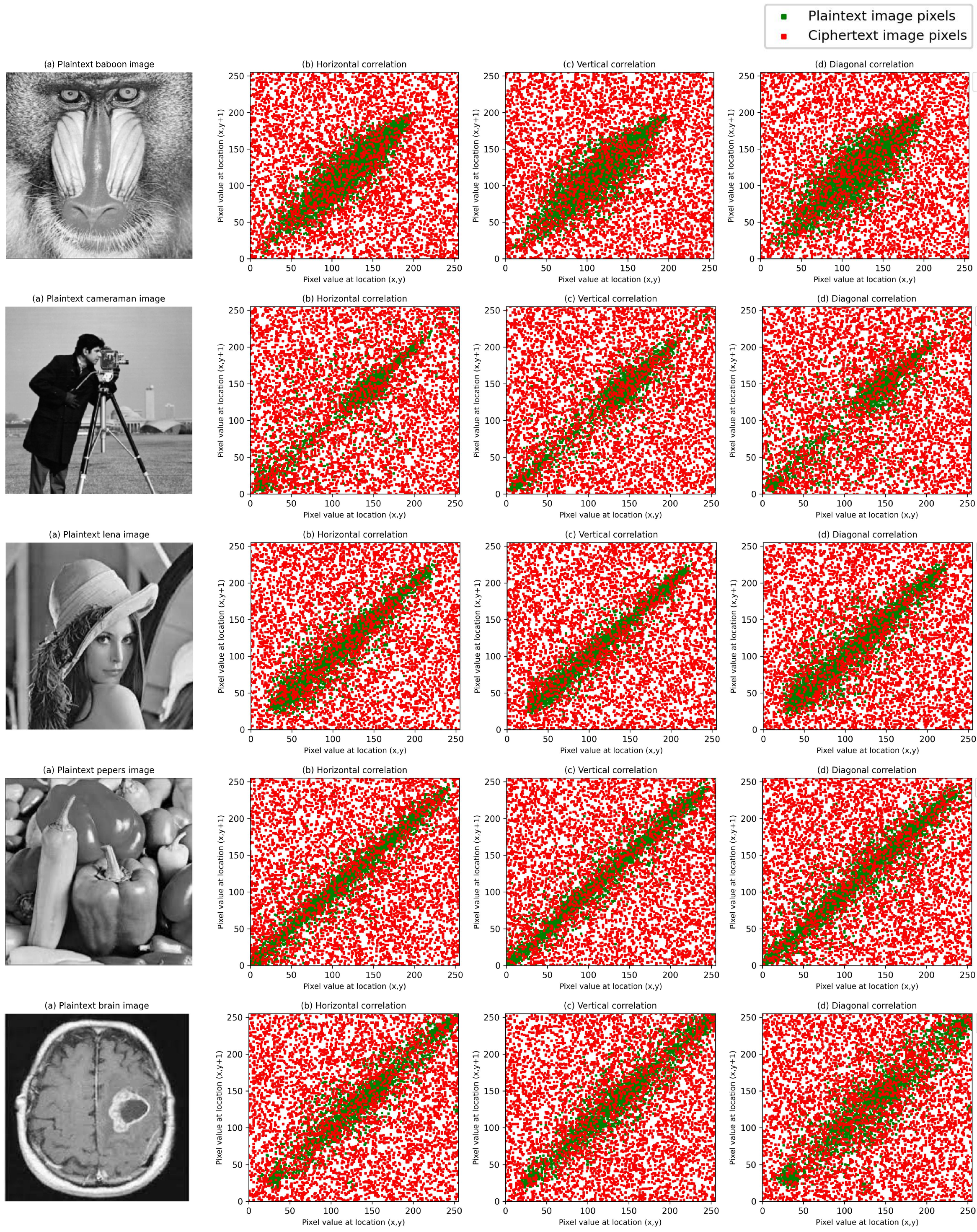}
    \caption{Correlation of different plaintext and their corresponding ciphertext images.}
    \label{fig:correlation_graphs}
\end{figure*}

To assess the effectiveness of the proposed scheme, 5,000 pairs of adjacent pixels were randomly sampled in horizontal, vertical, and diagonal directions from both the original and encrypted images. As shown in Figure~\ref{fig:correlation_graphs}, the scatter plots reveal a significant reduction in correlation after encryption. Table~\ref{tab:correlation-2d-shifting} further supports this observation by quantifying the correlation values before and after the chaotic permutation stage, highlighting its ability to disrupt spatial redundancy and enhance security.

As seen in the results, the modified AES-128 algorithm significantly reduces the correlation in the encrypted images, confirming its ability to destroy the inherent spatial redundancy.

Additionally, Table~\ref{tab:Correlation analysis of different encryption schemes} presents a comparison of the correlation performance with the suggested encryption method and recent encryption methods.  The suggested algorithm outperforms the techniques outlined in~\cite{zhu, ping, diab, pak, ye_huang}, achieving the lowest correlation values for the images in horizontal, vertical, and diagonal directions.

\subsection{ Entropy Analysis}
Analyzing the distribution of pixel intensities is a widely used method to evaluate the randomness of encrypted images. Information entropy, in particular, serves as a key metric for quantifying this randomness. For an ideally encrypted 8-bit grayscale image, the entropy value should approach 8, indicating uniformly distributed pixel values. Lower values suggest the presence of patterns or predictability that could be exploited in statistical attacks~\cite{shannon1948mathematical, wu2011image}.

In this study, we computed the information entropy for several standard test images using Equation~\eqref{eq:entropy}:

\begin{equation}
\label{eq:entropy}
\text{Entropy}(I) = - \sum_{s=0}^{255} P(I_s) \log_2 P(I_s)
\end{equation}

Here, \( I_s \) denotes the grayscale intensity level, and \( P(I_s) \) is the probability of occurrence of that level in the image. The summation spans all 256 possible gray values.

Table~\ref{tab:security_analysis} shows the entropy values achieved by the proposed encryption scheme on multiple benchmark images, consistently approaching the theoretical maximum of 8.0. This reflects a high degree of randomness and confirms the effectiveness of the proposed method in concealing statistical features of the original image. Furthermore, Table~\ref{tab:Entropy analysis of encrypted images.} compares our results with other contemporary encryption methods. The comparison highlights the superior performance of our scheme in maintaining entropy, reinforcing its robustness against entropy-based statistical attacks. These findings demonstrate that the proposed algorithm generates ciphertexts nearly indistinguishable from purely random data, ensuring strong protection against information leakage through statistical means.

\subsection{Homogeneity Analysis}
To assess the texture randomness and internal spatial correlation within the encrypted image, a gray-level co-occurrence matrix (GLCM) \cite{glcm}-based homogeneity analysis was performed. The GLCM captures the frequency of pixel intensity pairs occurring at a defined spatial offset, providing insight into the local texture structure of an image.

For this experiment, the encrypted image was first quantized to 8 gray levels to improve statistical stability. A horizontal pixel offset $(dx=1, dy=0)$ was used to construct the GLCM \cite{glcm}, focusing on the correlation between neighboring pixel intensities. The homogeneity metric, defined in Eq.~\ref{eq:homogeneity}, was computed from the normalized GLCM:

\begin{equation}
\text{Homogeneity} = \sum_{A=0}^{L-1} \sum_{B=0}^{L-1} \frac{P(A,B)}{1 + |A - B|}
\label{eq:homogeneity}
\end{equation}

where $L$ is the number of quantization levels and $P(A,B)$ is the likelihood that gray levels $A$ and $B$ in the image will co-occur. More efficient encryption is indicated by a lower homogeneity measure.  The results shown in Table~\ref{tab:homogeneity-analysis} indicate that the suggested encryption system is more effective than current techniques, especially when it comes to statistical security metrics and cryptographic robustness.  Furthermore, the homogeneity values of many encrypted pictures processed with the suggested scheme are reported in Table~\ref{tab:homogeneity-analysis-doc}.  The scheme's ability to get rid of unnecessary patterns and strengthen defenses against statistical attacks is further supported by the low homogeneity values, which validate the disruption of local pixel similarity.

\subsection{Energy Analysis}

An image's energy level corresponds to how much information it contains; images with more information have higher energy levels.  For image encryption to be highly secure, the encryption technique must create ciphertext images with as little visible information as possible.  According to mathematics, an image's energy is equal to the sum of the squared components of its Gray-Level Co-occurrence Matrix (GLCM):

\begin{equation}
\text{Energy} = \sum_{A=0}^{255} \sum_{B=0}^{255} \left[ c(A,B) \right]^2
\end{equation}

where \(c(A,B)\) indicates the encrypted image's GLCM \cite{glcm}.  The energy values of plaintext photos, a current encryption system, and the suggested encryption scheme are compared in Table~\ref{tab:energy-analysis}.  According to the results, the suggested approach produces encrypted images with lower energy values, indicating a higher level of unpredictability and enhanced encryption efficacy.  Furthermore, the energy values calculated for a number of encrypted images produced exclusively by the suggested technique are shown in Table~\ref{tab:energy-analysis-doc}, confirming its capacity to reduce pixel intensity regularity and hide structural features.

\subsection{Lossless Analysis}

The effectiveness of the proposed image encryption scheme is assessed using two widely accepted statistical metrics: Mean Squared Error (MSE) and Peak Signal-to-Noise Ratio (PSNR)~\cite{psnr_ssim}. These metrics are traditionally used to quantify distortion between two images and are adapted here to measure how effectively the encryption process obscures the original content.

PSNR calculates the ratio between the maximum possible power of a signal and the power of corrupting noise that affects its representation. In image encryption, a low PSNR value is desirable, as it indicates a significant difference between the plaintext and the ciphertext images. Conversely, MSE evaluates the average squared intensity difference between corresponding pixels of the original and encrypted images. A higher MSE value corresponds to greater distortion, which is favorable for encryption strength.

The formulas for these metrics are defined as follows:

\begin{equation}
\text{MSE} = \frac{1}{AB} \sum_{i=0}^{A-1} \sum_{j=0}^{B-1} \left[P(i,j) - E(i,j)\right]^2
\label{eq:mse}
\end{equation}

\begin{equation}
\text{PSNR} = 10\log_{10} \left( \frac{M_{\text{max}}^2}{\text{MSE}} \right)
\label{eq:psnr}
\end{equation}

where \(A \times B\) represents the image size, \(P(i,j)\) and \(E(i,j)\) denote the pixel values of the original and encrypted images, respectively, and \(M_{\text{max}}\) is the maximum possible pixel intensity (typically 255 for 8-bit grayscale images).

In our experiments, the proposed encryption scheme consistently produced low PSNR values and high MSE values across several benchmark images. This confirms the algorithm’s capability to significantly alter the original image content, thereby enhancing its resistance to visual and statistical attacks. These findings support the scheme’s suitability for confidentiality-critical applications where data privacy is paramount.

\subsection{Avalanche Effect Analysis}

\begin{figure}[htbp]
    \centering
    \includegraphics[width=0.9\linewidth]{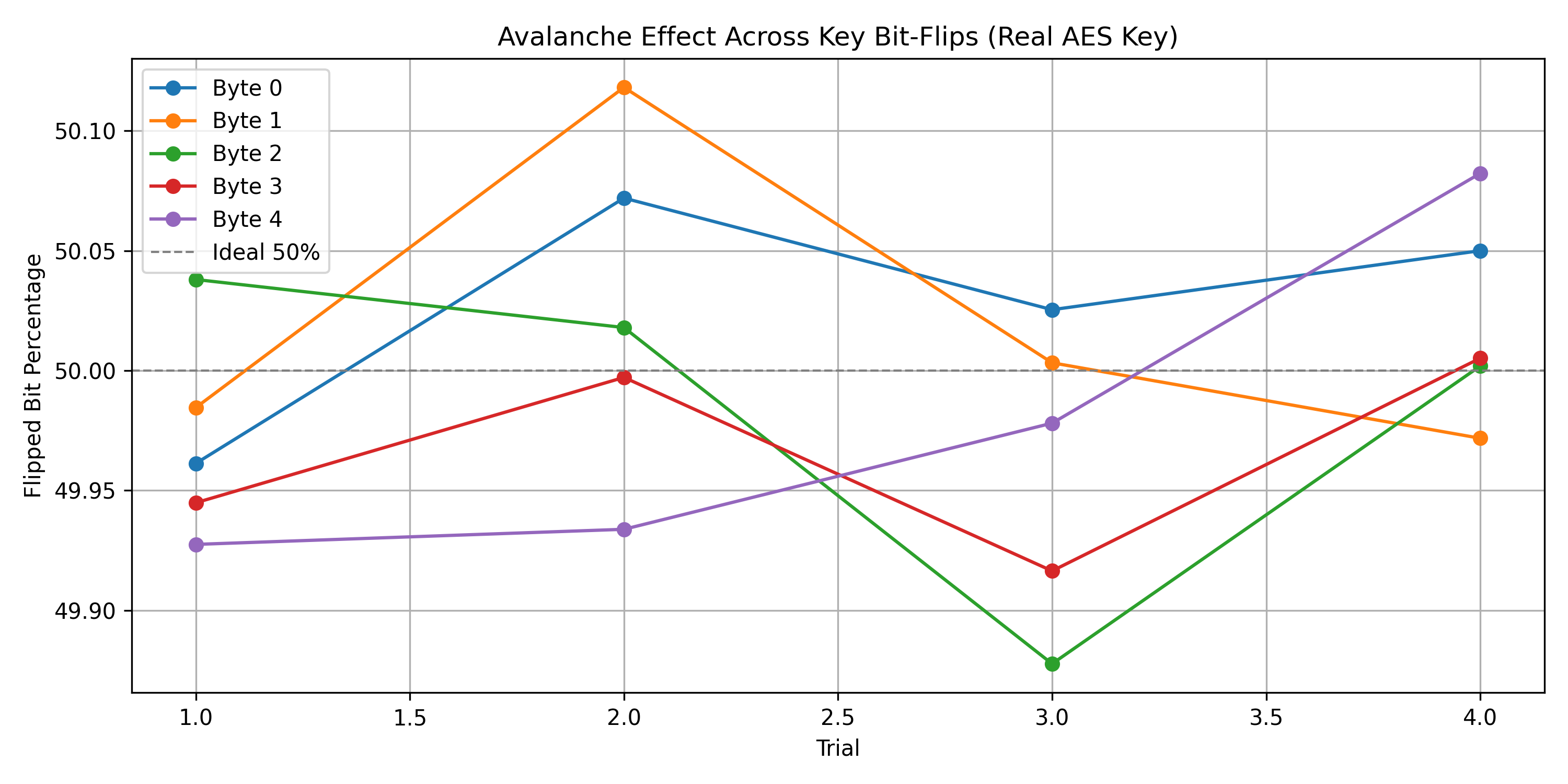}
    \caption{Avalanche effect across multiple key bit-flips.}
    \label{fig:avalanche-plot}
\end{figure}

The avalanche effect illustrates how a minor alteration to the input, like changing a single bit in the key, results in considerable modifications to the ciphertext.  Under such perturbations, a strong encryption method should ideally provide an output bit difference of 50\%.  We evaluated the bit changes in the ciphertext of a $256 \times 256$ grayscale image after flipping a single bit in different locations of the 128-bit AES key in order to assess this property.  Each key byte was tested four times at five different byte locations.  The flipped bit percentages stay closely concentrated around 50\%, with changes of less than 0.1\%, as seen in Figure~\ref{fig:avalanche-plot}.  This strengthens our scheme's defense against differential and brute-force attacks by confirming its great diffusion and key sensitivity.

\subsection{Differential Analysis}

Differential analysis is a critical technique for evaluating the robustness of image encryption algorithms. It assesses how sensitive the encryption scheme is to minor changes in the input image, such as altering a single pixel. Two widely recognized metrics for this purpose are the \textit{Number of Pixels Change Rate (NPCR)} and the \textit{Unified Average Changing Intensity (UACI)}~\cite{npcr_uachi}. These metrics help determine the algorithm’s resistance to differential attacks by quantifying the extent to which ciphertexts change when plaintexts are slightly altered.

NPCR measures the percentage of pixels that differ between two encrypted images—one generated from the original image and the other from a slightly modified version (e.g., with a single pixel changed). It is defined as:

\begin{equation}
\text{NPCR} = \frac{\sum_{i,j} D(i,j)}{A \times B} \times 100\%
\label{eq:npcr}
\end{equation}

where
\[
D(i,j) = 
\begin{cases}
0, & \text{if } E_1(i,j) = E_2(i,j) \\
1, & \text{if } E_1(i,j) \neq E_2(i,j)
\end{cases}
\]

UACI, on the other hand, calculates the average intensity difference between two cipher images and is given by:

\begin{equation}
\text{UACI} = \frac{1}{A \times B} \sum_{i,j} \left( \frac{|E_1(i,j) - E_2(i,j)|}{255} \right) \times 100\%
\label{eq:uaci}
\end{equation}

In these equations, \(E_1\) and \(E_2\) denote the two ciphertext images, and \(A \times B\) is the image resolution.

As shown in Table~\ref{tab:npcr-analysis} and Table~\ref{tab:uaci-analysis}, the proposed encryption scheme demonstrates superior NPCR and UACI values compared to existing approaches. High NPCR values indicate that the ciphertext undergoes significant changes even with minimal alterations in the plaintext, while high UACI values suggest strong diffusion of intensity changes. Table~\ref{tab:differential_attack} further confirms the scheme’s effectiveness by presenting consistent NPCR and UACI results across multiple encrypted test images. These outcomes validate the proposed algorithm’s robustness against differential attacks and its ability to maintain high security standards.

\subsection{ Key Space}
An image encryption algorithm's key space should be resistant to multiple attacks. particularly brute-force attacks. According to cryptographic standards, a secure encryption system should have a key space larger than $2^{128}$~\cite{light_weight_aes}. The proposed encryption algorithm integrates AES-128 with several key-dependent chaotic parameters. Specifically, it utilizes logistic map parameters $x_0$ and $r$, each with a computational precision of $10^{-15}$, Hénon map initial values $x_0$ and $y_0$, each with a precision of $10^{-15}$, a 128-bit initialization vector (IV) and a conventional 128-bit AES key.  With this configuration, the chaotic parameters' contribution to the key space is computed as follows:
\[
10^{15} \times 10^{15} \times 10^{15} \times 10^{15} = 10^{60} \approx 2^{199.2}
\]
Including the AES key and IV, the total key space becomes:
\[
2^{199.2} \times 2^{128} \times 2^{128} = 2^{455.2}
\]
Thus, the effective key length of the proposed system is approximately 455 bits, which significantly exceeds the key space of classical AES-128 ($2^{128}$) and AES-256 ($2^{256}$), making the proposed system highly resistant to brute-force attacks.

\subsection{Key Sensitivity Analysis}

An essential characteristic of a robust encryption scheme is its sensitivity to the secret key. A highly secure system must ensure that even a minor change in the encryption key, such as altering a single byte—produces a drastically different ciphertext. To evaluate this property, a key sensitivity experiment was conducted. Specifically, a 128-bit AES key was modified by just one byte, and the resulting ciphertext was decrypted using this incorrect key.

To quantify the structural differences between the correctly and incorrectly decrypted images, the Structural Similarity Index (SSIM) was employed. As shown in Table~\ref{tab:key_sensitivity}, the SSIM values for all test images were extremely low, ranging from 0.0012 for the \textit{FingerPrint} image to 0.0115 for the \textit{Surveillance} image. These values indicate minimal structural resemblance between the original and incorrectly decrypted outputs.

The results confirm that the proposed encryption scheme exhibits high key sensitivity: even a one-byte key variation leads to complete decryption failure. This behavior significantly strengthens the system's resilience against brute-force and key-related differential attacks, ensuring that unauthorized access is effectively thwarted.

\subsection{Cropping Attack Analysis}
\begin{figure}[htbp]
    \centering
    \includegraphics[width=0.48\textwidth]{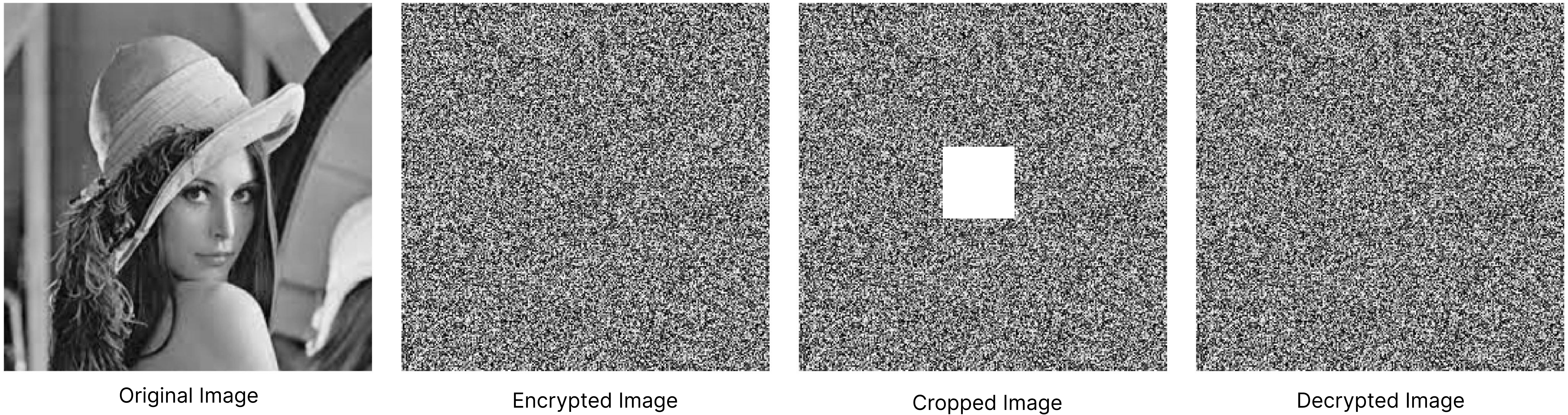}
    \caption{Cropping attack analysis.}
    \label{fig:crop_attack}
\end{figure}

A cropping attack scenario was simulated in order to assess the integrity and resilience of the suggested encryption system against localized ciphertext modification or leakage. A $50 \times 50$ region was extracted from the encrypted image in Figure\ref{fig:crop_attack} and independently decrypted. As shown in Figure\ref{fig:crop_attack}, the decrypted image --- with the cropped region reinserted --- does not yield any perceptible visual information. Despite applying the correct decryption key and parameters, the localized decryption of cropped ciphertext produces only randomized noise. This confirms that the proposed encryption framework, due to its block chaining, dynamic S-box substitution, and feedback diffusion mechanisms, resists partial decryption and offers strong protection against cropping-based attacks.

\subsection{Data Loss Resilience Analysis}
\begin{figure}[htbp]
    \centering
    \includegraphics[width=0.48\textwidth]{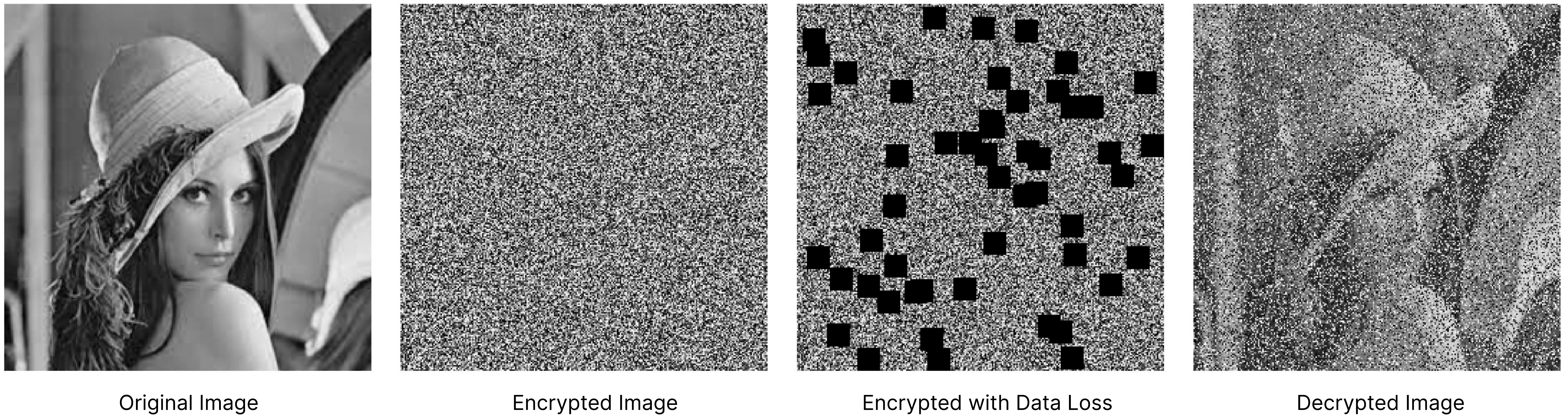}
    \caption{Data Loss attack analysis.}
    \label{fig:dataloss_attack}
\end{figure}

An experiment with block-wise data corruption was carried out to evaluate the durability of the suggested encryption technique against partial ciphertext loss. Specifically, a 20\% data loss was simulated by replacing randomly selected $16 \times 16$ blocks in the encrypted image with zero-valued blocks. The corrupted ciphertext was then decrypted using the correct key and configuration. As shown in Figure~\ref{fig:dataloss_attack}, the resulting decrypted image exhibits severe visual distortion and random artifacts, effectively preventing any meaningful reconstruction of the original content. This confirms the scheme’s high sensitivity to ciphertext integrity, making it resistant to partial decryption and data loss-based attacks. Quantitative evaluation over 10 iterations, as presented in Table~\ref{tab:data_loss_updated}, further supports these findings, where the average SSIM and PSNR values remain extremely low, highlighting the encryption system’s strong dependence on complete ciphertext integrity for successful decryption.

\subsection{Noise Attack Analysis}
\begin{figure}[htbp]
    \centering
    \includegraphics[width=0.48\textwidth]{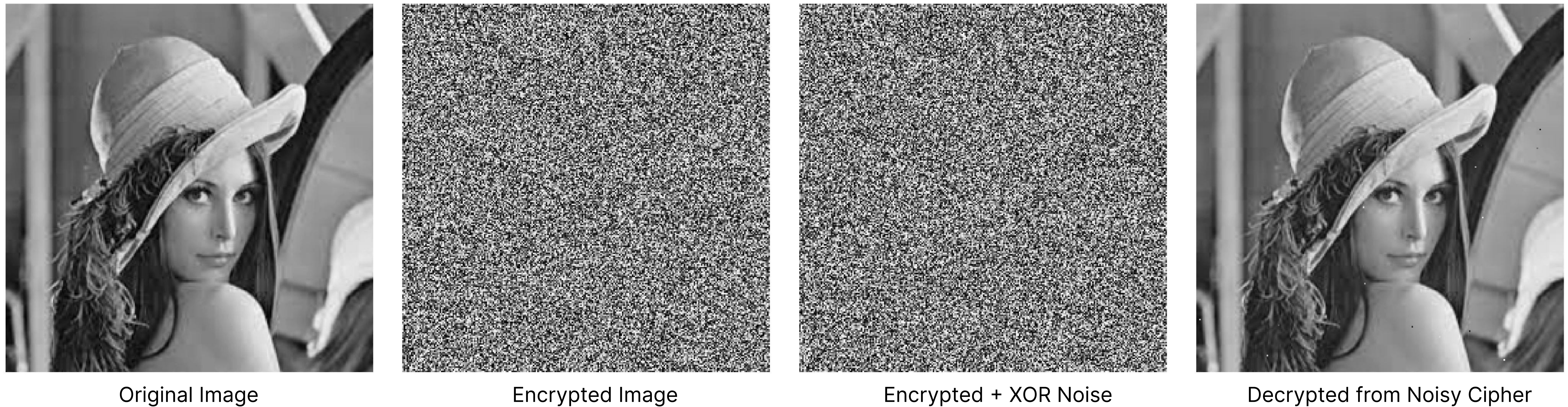}
    \caption{XOR-based noise attack analysis.}
    \label{fig:xor_noise_attack}
\end{figure}

In order to assess the suggested encryption scheme's resilience in noisy settings, multiple noise attacks were simulated on the ciphertext. First, a uniform XOR-based perturbation was applied by masking each ciphertext byte with a fixed binary pattern ($00001110$), mimicking low-level tampering or transmission errors. As illustrated in Figure~\ref{fig:xor_noise_attack}, the decrypted image remained visually close to the original, with only minor distortions, indicating strong fault tolerance under bit-level corruption. In addition, more realistic noise models were tested. Salt-and-pepper noise and Gaussian noise were injected into the ciphertext image to simulate impulsive and continuous random interference, respectively. The decrypted outputs maintained acceptable perceptual quality despite the presence of noise. Quantitative results for Salt-and-pepper noise and Gaussian noise are presented in Table \ref{tab:salt_pepper_noise} and Table \ref{tab:gaussian_noise} confirm this observation, with structural similarity (SSIM) and peak signal-to-noise ratio (PSNR) values demonstrating that the proposed scheme maintains visual coherence and encryption integrity under practical noise conditions.

\section{Conclusion}

This paper introduced a chaos-enhanced AES image encryption scheme with a secure QR-based dual-key distribution method. Addressing the limitations of classical AES and the lack of practical key-sharing solutions, our approach integrates dynamic S-boxes from a 2D Hénon map, logistic map-guided permutation, and feedback XOR diffusion to strengthen confusion and diffusion. Keys are covertly shared using a steganographically modified QR code, ensuring secure transmission. According to experimental data, the suggested system works better than current techniques, reaching near-ideal entropy ($\sim$7.997), high NPCR ($\sim$99.6\%), and UACI ($\sim$50.1\%), together with strong key sensitivity and minimal pixel correlation. These metrics confirm enhanced resistance to statistical and differential attacks. Overall, the method ensures robust, covert, and efficient image protection, making it suitable for sensitive domains like surveillance and medical imaging. The current implementation is restricted to grayscale images of fixed size ($256 \times 256$ pixels). RGB images and images of varying dimensions were not evaluated in this study. Additionally, while the QR-based key distribution scheme proved effective under controlled conditions, its resilience to image compression, distortion, or loss during transmission was not assessed. These constraints limit the method’s direct applicability in dynamic, real-world environments. We plan to extend the proposed system to support color (RGB) images and arbitrary resolutions. Another direction includes developing adaptive preprocessing steps for high-resolution or streaming data. Finally, integrating quantum-resistant public-key algorithms and optimizing performance for real-time encryption and decryption will be critical for deployment in emerging post-quantum security contexts.

% COMPLETE----------------------------COMPLETE-----------------------------------

% \section*{CRediT authorship contribution statement}
% \textbf{Araf Nishan:} Conceptualization, Methodology, Writing – original draft. \textbf{S. M. Taslim Uddin Raju:}  Conceptualization, Methodology, Writing – review \& editing. \textbf{Md Imran Hossain:} Visualization, Investigation. \textbf{Safin Ahmed Dipto:} Formal analysis, Visualization, Writing – original draft. \textbf{S. M. Tanvir Uddin:} Writing – review \& editing, Project administration. \textbf{Asif Sijan:} Investigation, Validation. \textbf{Md Abu Shahid Chowdhury:} Writing – review \& editing.  \textbf{Ashfaq Ahmad:} Writing – review \& editing, Visualization. \textbf{Md Mahamudul Hasan Khan:} Supervision.

% \section*{Transparency}

\section*{Declaration of Funding}
This paper was not funded. 

% \section*{Declaration of financial/other relationships}
% None. There is no any financial/other relationship. Peer reviewers on this manuscript have no relevant financial or other relationships to disclose.

\section*{Author Contributions}
\textbf{Md Rishadul Bayesh:} Conceptualization, Methodology, Formal analysis, Visualization, Writing – original draft. \textbf{Dabbrata Das:}  Conceptualization, Formal analysis, Writing – original draft \& review, Project administration. \textbf{Md Ahadullah:} Formal analysis, Investigation and Review.

\section*{Ethical Approval}
Not required.

\section*{Declaration of Competing Interest}
The authors declare no competing interests.

\section*{Acknowledgements}
This work is supported in part by the Uttara University.
% \end{flushleft}



\bibliographystyle{cas-model2-names}
% \bibliographystyle{cas-model2-names}

\bibliography{refs}

\end{document}